
\magnification=\magstep1

\def\spa#1.#2{\left\langle#1\,#2\right\rangle}
\def\spb#1.#2{\left[#1\,#2\right]}
\newbox\SlashedBox
\def\slashed#1{\setbox\SlashedBox=\hbox{#1}
\hbox to 0pt{\hbox to 1\wd\SlashedBox{\hfil/\hfil}\hss}#1}
\def\hboxtosizeof#1#2{\setbox\SlashedBox=\hbox{#1}
\hbox to 1\wd\SlashedBox{#2}}

\def\ifsmall{\iffalse}  
\def\titlepagefont{}  

\def\DefineTeXgraphics{%
\special{ps::[global] /TeXgraphics { } def}}  

\def\today{\ifcase\month\or January\or February\or March\or April\or May
\or June\or July\or August\or September\or October\or November\or
December\fi\space\number\day, \number\year}
\def\eatPrefix19{}
\def\Year{\expandafter\eatPrefix\the\year}
\newcount\hours \newcount\minutes
\def\monthname{\ifcase\month\or
January\or February\or March\or April\or May\or June\or July\or
August\or September\or October\or November\or December\fi}
\def\shortmonthname{\ifcase\month\or
Jan\or Feb\or Mar\or Apr\or May\or Jun\or Jul\or
Aug\or Sep\or Oct\or Nov\or Dec\fi}

\def\TimeStamp{\hours\the\time\divide\hours by60%
\minutes -\the\time\divide\minutes by60\multiply\minutes by60%
\advance\minutes by\the\time%
${\rm \shortmonthname}\cdot\if\day<10{}0\fi\the\day\cdot\the\year%
\qquad\the\hours:\if\minutes<10{}0\fi\the\minutes$}







\newif\ifdraftmode
\newif\ifleftlabels

\def\nolabels{\def\wrlabeL##1{}\def\eqlabeL##1{}\def\reflabeL##1{}}
\def\writelabels{\def\wrlabeL##1{\leavevmode\vadjust{\rlap{\smash%
{\line{{\escapechar=` \hfill\rlap{\sevenrm\hskip.03in\string##1}}}}}}}%
\def\eqlabeL##1{{\escapechar-1\rlap{\sevenrm\hskip.05in\string##1}}}%
\def\reflabeL##1{\noexpand\rlap{\noexpand\sevenrm[\string##1]}}}
\def\writeleftlabels{\def\wrlabeL##1{\leavevmode\vadjust{\rlap{\smash%
{\line{{\escapechar=` \hfill\rlap{\sevenrm\hskip.03in\string##1}}}}}}}%
\def\eqlabeL##1{{\escapechar-1%
\rlap{\sixrm\hskip.05in\string##1}%
\llap{\sevenrm\string##1\hskip.03in\hbox to \hsize{}}}}%
\def\reflabeL##1{\noexpand\rlap{\noexpand\sevenrm[\string##1]}}}
\nolabels

\newdimen\fullhsize
\newdimen\hstitle
\hstitle=\hsize 
\newdimen\hsbody
\hsbody=\hsize 
\newdimen\hbodyoffset
\hbodyoffset=\hoffset 
\newbox\leftpage
\def\abstract#1{#1}
\def\rotated{\special{ps: landscape}
\magnification=1000  
\baselineskip=14pt
\global\hstitle=9truein\global\hsbody=4.75truein
\global\vsize=7truein\global\voffset=-.31truein
\global\hoffset=-0.54in\global\hbodyoffset=-.54truein
\global\fullhsize=10truein
\def\DefineTeXgraphics{%
\special{ps::[global]
/TeXgraphics {currentpoint translate 0.7 0.7 scale
              -80 0.72 mul -1000 0.72 mul translate} def}}
\let\lr=L
\def\ifsmall{\iftrue}
\def\titlepagefont{\twelvepoint}
\trueseventeenpoint
\def\almostshipout##1{\if L\lr \count1=1
      \global\setbox\leftpage=##1 \global\let\lr=R
   \else \count1=2
      \shipout\vbox{\hbox to\fullhsize{\box\leftpage\hfil##1}}
      \global\let\lr=L\fi}

\output={\ifnum\count0=1 
 \shipout\vbox{\hbox to \fullhsize{\hfill\pagebody\hfill}}\advancepageno
 \else
 \almostshipout{\leftline{\vbox{\pagebody\makefootline}}}\advancepageno
 \fi}

\def\abstract##1{{\leftskip=1.5in\rightskip=1.5in ##1\par}} }

\def\linemessage#1{\immediate\write16{#1}}

\global\newcount\secno \global\secno=0
\global\newcount\appno \global\appno=0
\global\newcount\meqno \global\meqno=1
\global\newcount\subsecno \global\subsecno=0
\global\newcount\figno \global\figno=0

\newif\ifAnyCounterChanged
\let\terminator=\relax
\def\normalize#1{\ifx#1\terminator\let\next=\relax\else%
\if#1i\aftergroup i\else\if#1v\aftergroup v\else\if#1x\aftergroup x%
\else\if#1l\aftergroup l\else\if#1c\aftergroup c\else%
\if#1m\aftergroup m\else%
\if#1I\aftergroup I\else\if#1V\aftergroup V\else\if#1X\aftergroup X%
\else\if#1L\aftergroup L\else\if#1C\aftergroup C\else%
\if#1M\aftergroup M\else\aftergroup#1
\fi\fi\fi\fi\fi\fi\fi\fi\fi\fi\fi\fi%
\let\next=\normalize\fi%
\next}
\def\makeNormal#1#2{\def\doNormalDef{\edef#1}\begingroup%
\aftergroup\doNormalDef\aftergroup{\normalize#2\terminator\aftergroup}%
\endgroup}

\def\warnIfChanged#1#2{%
\ifundef#1
\else\begingroup%
\edef\oldDefinitionOfCounter{#1}\edef\newDefinitionOfCounter{#2}%
\ifx\oldDefinitionOfCounter\newDefinitionOfCounter%
\else%
\linemessage{Warning: definition of \noexpand#1 has changed.}%
\global\AnyCounterChangedtrue\fi\endgroup\fi}

\def\Section#1{\global\advance\secno by1\relax\global\meqno=1%
\global\subsecno=0%
\bigbreak\bigskip
\centerline{\twelvepoint \bf %
\the\secno. #1}%
\par\nobreak\medskip\nobreak}
\def\tagsection#1{%
\warnIfChanged#1{\the\secno}%
\xdef#1{\the\secno}%
\ifWritingAuxFile\immediate
\write\auxfile{\noexpand\xdef\noexpand#1{#1}}\fi%
}

\def\subsection{\Subsection}

\def\romappno{\uppercase\expandafter{\romannumeral\appno}}
\def\makeNormalizedRomappno{%
\expandafter\makeNormal\expandafter\normalizedromappno%
\expandafter{\romannumeral\appno}%
\edef\normalizedromappno{\uppercase{\normalizedromappno}}}
\def\Appendix#1{\global\advance
\appno by1\relax\global\meqno=1\global\secno=0
\bigbreak\bigskip
\centerline{\twelvepoint \bf Appendix %
\romappno. #1}%
\par\nobreak\medskip\nobreak}
\def\tagappendix#1{\makeNormalizedRomappno%
\warnIfChanged#1{\normalizedromappno}%
\xdef#1{\normalizedromappno}%
\ifWritingAuxFile\immediate
\write\auxfile{\noexpand\xdef\noexpand#1{#1}}\fi%
}
\def\appendix{\Appendix}

\def\eqn#1{\makeNormalizedRomappno%
\ifnum\secno>0%
  \warnIfChanged#1{\the\secno.\the\meqno}%
  \eqno(\the\secno.\the\meqno)\xdef#1{\the\secno.\the\meqno}%
     \global\advance\meqno by1
\else\ifnum\appno>0%
  \warnIfChanged#1{\normalizedromappno.\the\meqno}%
  \eqno({\rm\romappno}.\the\meqno)%
      \xdef#1{\normalizedromappno.\the\meqno}%
     \global\advance\meqno by1
\else%
  \warnIfChanged#1{\the\meqno}%
  \eqno(\the\meqno)\xdef#1{\the\meqno}%
     \global\advance\meqno by1
\fi\fi%
\eqlabeL#1%
\ifWritingAuxFile\immediate\write\auxfile{
\noexpand\xdef\noexpand#1{#1}}\fi%
}
\def\defeqn#1{\makeNormalizedRomappno%
\ifnum\secno>0%
  \warnIfChanged#1{\the\secno.\the\meqno}%
  \xdef#1{\the\secno.\the\meqno}%
     \global\advance\meqno by1
\else\ifnum\appno>0%
  \warnIfChanged#1{\normalizedromappno.\the\meqno}%
  \xdef#1{\normalizedromappno.\the\meqno}%
     \global\advance\meqno by1
\else%
  \warnIfChanged#1{\the\meqno}%
  \xdef#1{\the\meqno}%
     \global\advance\meqno by1
\fi\fi%
\eqlabeL#1%
\ifWritingAuxFile\immediate\write\auxfile{\noexpand\xdef
\noexpand#1{#1}}\fi%
}
\def\anoneqn{\makeNormalizedRomappno%
\ifnum\secno>0
  \eqno(\the\secno.\the\meqno)%
     \global\advance\meqno by1
\else\ifnum\appno>0
  \eqno({\rm\normalizedromappno}.\the\meqno)%
     \global\advance\meqno by1
\else
  \eqno(\the\meqno)%
     \global\advance\meqno by1
\fi\fi%
}
\def\mfig#1#2{\global\advance\figno by1%
\relax#1\the\figno%
\warnIfChanged#2{\the\figno}%
\edef#2{\the\figno}%
\reflabeL#2%
\ifWritingAuxFile\immediate\write\auxfile{
\noexpand\xdef\noexpand#2{#2}}\fi%
}

\catcode`@=11 

\font\titlefont=cmr10 at 16pt
\font\ninerm=cmr9
\font\eightrm=cmr8
\font\sixrm=cmr6

\def\loadtrueseventeenpoint{
 \font\seventeenrm=cmr10 at 17.28truept
 \font\seventeeni=cmmi10 at 17.28truept
 \font\seventeenbf=cmbx10 at 17.28truept
 \font\seventeenit=cmti10 at 17.28truept
 \font\seventeensl=cmsl10 at 17.28truept
 \font\seventeensy=cmsy10 at 17.28truept
}
\def\loadfourteenpoint{
\font\fourteenrm=cmr10 at 14.4pt
\font\fourteeni=cmmi10 at 14.4pt
\font\fourteenit=cmti10 at 14.4pt
\font\fourteensl=cmsl10 at 14.4pt
\font\fourteensy=cmsy10 at 14.4pt
\font\fourteenbf=cmbx10 at 14.4pt
}
\def\loadtruetwelvepoint{
\font\twelverm=cmr10 at 12truept
\font\twelvei=cmmi10 at 12truept
\font\twelveit=cmti10 at 12truept
\font\twelvesl=cmsl10 at 12truept
\font\twelvesy=cmsy10 at 12truept
\font\twelvebf=cmbx10 at 12truept
}

\font\ninei=cmmi9
\font\eighti=cmmi8
\font\sixi=cmmi6
\skewchar\ninei='177 \skewchar\eighti='177 \skewchar\sixi='177

\font\ninesy=cmsy9
\font\eightsy=cmsy8
\font\sixsy=cmsy6
\skewchar\ninesy='60 \skewchar\eightsy='60 \skewchar\sixsy='60

\font\ninebf=cmbx9
\font\eightbf=cmbx8
\font\sixbf=cmbx6

\font\ninett=cmtt9
\font\eighttt=cmtt8

\hyphenchar\tentt=-1 
\hyphenchar\ninett=-1
\hyphenchar\eighttt=-1

\font\ninesl=cmsl9
\font\eightsl=cmsl8

\font\nineit=cmti9
\font\eightit=cmti8


\newskip\ttglue
\def\tenpoint{\def\rm{\fam0\tenrm}%
  \textfont0=\tenrm \scriptfont0=\sevenrm \scriptscriptfont0=\fiverm
  \textfont1=\teni \scriptfont1=\seveni \scriptscriptfont1=\fivei
  \textfont2=\tensy \scriptfont2=\sevensy \scriptscriptfont2=\fivesy
  \textfont3=\tenex \scriptfont3=\tenex \scriptscriptfont3=\tenex
  \def\it{\fam\itfam\tenit}\textfont\itfam=\tenit
  \def\sl{\fam\slfam\tensl}\textfont\slfam=\tensl
  \def\bf{\fam\bffam\tenbf}\textfont\bffam=\tenbf
  \scriptfont\bffam=\sevenbf
  \scriptscriptfont\bffam=\fivebf
  \normalbaselineskip=12pt
  \let\sc=\eightrm
  \let\big=\tenbig
  \setbox\strutbox=\hbox{\vrule height8.5pt depth3.5pt width\z@}%
  \normalbaselines\rm}

\def\twelvepoint{\def\rm{\fam0\twelverm}%
  \textfont0=\twelverm \scriptfont0=\ninerm \scriptscriptfont0=\sevenrm
  \textfont1=\twelvei \scriptfont1=\ninei \scriptscriptfont1=\seveni
  \textfont2=\twelvesy \scriptfont2=\ninesy \scriptscriptfont2=\sevensy
  \textfont3=\tenex \scriptfont3=\tenex \scriptscriptfont3=\tenex
  \def\it{\fam\itfam\twelveit}\textfont\itfam=\twelveit
  \def\sl{\fam\slfam\twelvesl}\textfont\slfam=\twelvesl
  \def\bf{\fam\bffam\twelvebf}\textfont\bffam=\twelvebf
  \scriptfont\bffam=\ninebf
  \scriptscriptfont\bffam=\sevenbf
  \normalbaselineskip=12pt
  \let\sc=\eightrm
  \let\big=\tenbig
  \setbox\strutbox=\hbox{\vrule height8.5pt depth3.5pt width\z@}%
  \normalbaselines\rm}

\def\fourteenpoint{\def\rm{\fam0\fourteenrm}%
  \textfont0=\fourteenrm \scriptfont0=\tenrm \scriptscriptfont0=\sevenrm
  \textfont1=\fourteeni \scriptfont1=\teni \scriptscriptfont1=\seveni
  \textfont2=\fourteensy \scriptfont2=\tensy \scriptscriptfont2=\sevensy
  \textfont3=\tenex \scriptfont3=\tenex \scriptscriptfont3=\tenex
  \def\it{\fam\itfam\fourteenit}\textfont\itfam=\fourteenit
  \def\sl{\fam\slfam\fourteensl}\textfont\slfam=\fourteensl
  \def\bf{\fam\bffam\fourteenbf}\textfont\bffam=\fourteenbf%
  \scriptfont\bffam=\tenbf
  \scriptscriptfont\bffam=\sevenbf
  \normalbaselineskip=17pt
  \let\sc=\elevenrm
  \let\big=\tenbig
  \setbox\strutbox=\hbox{\vrule height8.5pt depth3.5pt width\z@}%
  \normalbaselines\rm}

\def\seventeenpoint{\def\rm{\fam0\seventeenrm}%
  \textfont0=\seventeenrm \scriptfont0=\fourteenrm
  \scriptscriptfont0=\tenrm
  \textfont1=\seventeeni \scriptfont1=\fourteeni \scriptscriptfont1=\teni
  \textfont2=\seventeensy \scriptfont2=\fourteensy
  \scriptscriptfont2=\tensy
  \textfont3=\tenex \scriptfont3=\tenex \scriptscriptfont3=\tenex
  \def\it{\fam\itfam\seventeenit}\textfont\itfam=\seventeenit
  \def\sl{\fam\slfam\seventeensl}\textfont\slfam=\seventeensl
  \def\bf{\fam\bffam\seventeenbf}\textfont\bffam=\seventeenbf%
  \scriptfont\bffam=\fourteenbf
  \scriptscriptfont\bffam=\twelvebf
  \normalbaselineskip=21pt
  \let\sc=\fourteenrm
  \let\big=\tenbig
  \setbox\strutbox=\hbox{\vrule height 12pt depth 6pt width\z@}%
  \normalbaselines\rm}

\def\ninepoint{\def\rm{\fam0\ninerm}%
  \textfont0=\ninerm \scriptfont0=\sixrm \scriptscriptfont0=\fiverm
  \textfont1=\ninei \scriptfont1=\sixi \scriptscriptfont1=\fivei
  \textfont2=\ninesy \scriptfont2=\sixsy \scriptscriptfont2=\fivesy
  \textfont3=\tenex \scriptfont3=\tenex \scriptscriptfont3=\tenex
  \def\it{\fam\itfam\nineit}\textfont\itfam=\nineit
  \def\sl{\fam\slfam\ninesl}\textfont\slfam=\ninesl
  \def\bf{\fam\bffam\ninebf}\textfont\bffam=\ninebf
  \scriptfont\bffam=\sixbf
  \scriptscriptfont\bffam=\fivebf
  \normalbaselineskip=11pt
  \let\sc=\sevenrm
  \let\big=\ninebig
  \setbox\strutbox=\hbox{\vrule height8pt depth3pt width\z@}%
  \normalbaselines\rm}

\def\eightpoint{\def\rm{\fam0\eightrm}%
  \textfont0=\eightrm \scriptfont0=\sixrm \scriptscriptfont0=\fiverm%
  \textfont1=\eighti \scriptfont1=\sixi \scriptscriptfont1=\fivei%
  \textfont2=\eightsy \scriptfont2=\sixsy \scriptscriptfont2=\fivesy%
  \textfont3=\tenex \scriptfont3=\tenex \scriptscriptfont3=\tenex%
  \def\it{\fam\itfam\eightit}\textfont\itfam=\eightit%
  \def\sl{\fam\slfam\eightsl}\textfont\slfam=\eightsl%
  \def\bf{\fam\bffam\eightbf}\textfont\bffam=
  \eightbf \scriptfont\bffam=\sixbf%
  \scriptscriptfont\bffam=\fivebf%
  \normalbaselineskip=9pt%
  \let\sc=\sixrm%
  \let\big=\eightbig%
  \setbox\strutbox=\hbox{\vrule height7pt depth2pt width\z@}%
  \normalbaselines\rm}

\def\tenbig#1{{\hbox{$\left#1\vbox to8.5pt{}\right.\n@space$}}}
\def\ninebig#1{{\hbox{$\textfont0=\tenrm\textfont2=\tensy
  \left#1\vbox to7.25pt{}\right.\n@space$}}}
\def\eightbig#1{{\hbox{$\textfont0=\ninerm\textfont2=\ninesy
  \left#1\vbox to6.5pt{}\right.\n@space$}}}

\def\footnote#1{\edef\@sf{\spacefactor\the\spacefactor}#1\@sf
      \insert\footins\bgroup\eightpoint
      \interlinepenalty100 \let\par=\endgraf
        \leftskip=\z@skip \rightskip=\z@skip
        \splittopskip=10pt plus 1pt minus 1pt \floatingpenalty=20000
        \smallskip\item{#1}\bgroup\strut\aftergroup\@foot\let\next}
\skip\footins=12pt plus 2pt minus 4pt
\dimen\footins=30pc 

\newinsert\margin
\dimen\margin=\maxdimen
\def\titlefont{\seventeenpoint}
\loadtruetwelvepoint 
\loadtrueseventeenpoint
\catcode`\@=\active
\catcode`@=12  
\catcode`\"=\active

\def\eatOne#1{}
\def\ifundef#1{\expandafter\ifx%
\csname\expandafter\eatOne\string#1\endcsname\relax}
\def\notTrue{\iffalse}\def\isTrue{\iftrue}
\def\ifdef#1{{\ifundef#1%
\aftergroup\notTrue\else\aftergroup\isTrue\fi}}

\global\newcount\refno \global\refno=1
\newwrite\rfile
\newlinechar=`\^^J
\def\ref#1#2{\the\refno\nref#1{#2}}
\def\nref#1#2{\xdef#1{\the\refno}%
\ifnum\refno=1\immediate\openout\rfile=refs.tmp\fi%
\immediate\write\rfile{\noexpand\item{[\noexpand#1]\ }#2.}%
\global\advance\refno by1}
\def\lref#1#2{\the\refno\xdef#1{\the\refno}%
\ifnum\refno=1\immediate\openout\rfile=refs.tmp\fi%
\immediate\write\rfile{\noexpand\item{[\noexpand#1]\ }#2\semi}%
\global\advance\refno by1}
\def\cref#1{\immediate\write\rfile{#1\semi}}
\def\eref#1{\immediate\write\rfile{#1.}}

\def\semi{;\hfil\noexpand\break}

\def\inputAuxIfPresent#1{\immediate\openin1=#1
\ifeof1\message{No file \auxfileName; I'll create one.
}\else\closein1\relax\input\auxfileName\fi%
}

\newif\ifWritingAuxFile
\newwrite\auxfile
\def\SetUpAuxFile{%
\xdef\auxfileName{\jobname.aux}%
\inputAuxIfPresent{\auxfileName}%
\WritingAuxFiletrue%
\immediate\openout\auxfile=\auxfileName}


\def\bye{\par\vfill\supereject%
\ifAnyCounterChanged\linemessage{
Some counters have changed.  Re-run tex to fix them up.}\fi%
\end}

\def\Ms{$M_s$}
\def\Split{\mathop{\rm Split}\nolimits}
\def\lamb{$\Lambda$}
\def\alphas{$\alpha_s$}
\def\gcub{${\rm tr}(G^3)$}

\overfullrule 0pt
\loadfourteenpoint
{\nopagenumbers\hsize=\hstitle\vskip1in
\hfuzz 35 pt
\vbadness=10001

\rightline{   }
\rightline{   }
\rightline{SLAC--PUB--6416}
\rightline{December 1993}
\rightline{(T)}
\rightline{   }

\vskip 0.5 in

\centerline{\titlefont{Testing Gluon Self-Interactions in}}
\centerline{\titlefont{Three Jet Events at Hadron Colliders${}^{\star}$}}
\vskip0.5in

\centerline{Lance Dixon and Yael Shadmi}

\vskip0.2in
\centerline{\it Stanford Linear Accelerator Center}
\centerline{\it Stanford, CA 94309}

\vskip 0.2in
\baselineskip15pt

\vskip 0.75truein
\centerline{\bf Abstract}

{
The effective operator ${\rm tr}(G^3)$
is the only dimension-6 gluonic operator that cannot be related to
four-quark operators. A peculiar property of this operator is
that it does not contribute to two-jet production at hadron colliders,
at the level of one operator insertion and
leading-order in $\alpha_s$; therefore we
study its effects on three jet events.
To calculate the helicity amplitudes induced by this operator
we make extensive use of collinear factorization.
We propose several ways of detecting the ${\rm tr}(G^3)$ signal,
one of which exploits its non-trivial behavior under azimuthal
rotations of two almost collinear jets.}

\vskip 0.5truein

\centerline{\sl Submitted to Nuclear Physics B}

\vfill
\vskip 0.7in
\noindent\hrule width 3.6in\hfil\break
${}^{\star}$Research supported by the Department of
Energy under grant DE-AC03-76SF00515.\hfil\break
%


\eject}

\baselineskip 15pt


\line{ {\sl 1. Introduction}\hfill }
\vskip0.1truein

The gluon self-coupling is perhaps the clearest manifestation of the
non-abelian nature of QCD. Tests of QCD are therefore incomplete
without quantitative tests of the gluonic sector. New physics which
respects the symmetries of QCD can produce deviations from the QCD
prediction for the gluon self-coupling at higher energies. The
deviations can be parametrized by an effective Lagrangian including
higher-dimension operators which describe the low-energy effects of the
new physics. By calculating the effect of these operators on partonic
scattering, one can set lower bounds on the characteristic energy
scale, \lamb, of the new physics in a manner independent of the details
of the new physics.

The use of higher dimension operators as probes for new physics was
originally suggested for interactions of leptons and
quarks~[\ref\peskin{E. Eichten, K. Lane, and M. Peskin, Phys. Rev.
Lett. 50 (1983) 811}].
In the quark case, the dimension-6 operators are
four-quark contact operators and they affect $2\to 2$ quark scattering
($q\bar{q}\to q\bar{q}$ etc.) at leading-order. Measurements of the
dijet invariant mass distribution at 1.8 TeV by the CDF collaboration
have recently led to a bound of 1.3 TeV on the scale
$\Lambda$  associated with such operators~[\ref \cdfq{CDF
Collaboration, F. Abe {\it et al.}, FERMILAB-PUB-93/151-E}], where
$\Lambda$ is defined by the conventions of ref.~[1].

This approach was first applied to testing purely gluonic
interactions by Simmons in ref.~[\ref\simmons{E.H. Simmons, Phys.
Lett. B226 (1989) 132; Phys. Lett. B246 (1990) 471}].
It turns out, for reasons
we will discuss in the following, that placing bounds on the scale of
deviations from QCD in the gluon sector is more difficult than in the
quark sector. It is this difficulty which the present work addresses.
Though most previous work has concentrated on the two-jet cross-section,
we will argue here that the gluonic dimension-6 operator
\gcub\ can be bounded most efficiently
by studying three jet events at hadron colliders.

To test the gluon sector of QCD quantitatively, an effective Lagrangian
that preserves the main features of QCD is desirable. At the level of
dimension-6 operators the unique effective Lagrangian that can be
constructed from gluon fields alone, and that is $SU(3)$ gauge-invariant
and CP-even is~[\simmons]
$$
{\cal L_{{\rm eff}}}\ =\ {\cal L_{{\rm QCD}}}\ +\
C{g_s\over{\Lambda}^2}f_{abc}G^{a\mu}_\nu G^{b\nu}_\rho
G^{c\rho}_\mu\ +\ C^\prime{1\over {\Lambda^\prime}^2}
D_{\mu} G^{a\mu}_\nu D_\rho G^{a\rho\nu}\ ,
\eqn\efflag
$$
where  $G_{\mu\nu}=-(i/g_s)[D_\mu,D_\nu]$ is the gluon field strength,
$D_\mu={\partial}_\mu+i g_s A_\mu$ is the covariant derivative,
$\Lambda$ and $\Lambda^\prime$ are the characteristic energy scales
of the new physics and $C$
and $C^\prime$ are numerical coefficients. In accord with the
conventions of ref.~[\peskin], we take $C = C^\prime = 4\pi$; then this
equation defines $\Lambda$ and $\Lambda^\prime$.

The new physics could be associated with heavy colored particles which
couple to gluons and affect their interactions through loops, in which
case the scales $\Lambda$, $\Lambda^\prime$
would be proportional to the mass of the heavy
particle, with a proportionality factor that depends on the details of
the new physics, such as the spin and color of the heavy particle.
Alternatively, the new physics could be gluon compositeness, in which
case $\Lambda$, $\Lambda^\prime$ would
be proportional to the scale of the interaction
between the gluon constituents. In this work we do not concern
ourselves with the possible source of new physics, but rather discuss
the implications of the effective Lagrangian~(\efflag) as a
model-independent test of gluon self-interactions.

Using the equation of motion for the gluon field $A_\mu$, the second
operator in~(\efflag) can be written as a four-quark contact
operator~[\simmons]. Thus, it cannot provide a clean test of the gluon
sector independently of quark effects. Also, since this operator is
equivalent to a four-quark operator, it is fairly easy to obtain a good
bound on the scale $\Lambda$ associated with it just as in the case of
four-quark operators;
Simmons and Cho have recently estimated this bound as 2.03
TeV~[\ref\simcho{P. Cho and E. Simmons,
preprint HUTP-93/A018, hep-ph/9307345}].

In the following we will therefore neglect the second operator
in~(\efflag) (i.e., we set $\Lambda^\prime = \infty$),
as well as four-quark operators, and focus on the
first operator in~(\efflag), to which we refer as \gcub.

Unfortunately, unlike the four-quark contact operators which interfere
with pure QCD at leading-order in $\alpha_s$ and in $1/\Lambda^2$, the
leading-order contribution of \gcub\ to massless-parton scattering
vanishes: the four-parton tree amplitudes containing one insertion of
this operator have a helicity structure which is orthogonal to that of
four-parton QCD tree amplitudes when the four partons are
massless~[\simmons]. A leading order contribution from \gcub\ does
appear in four jet decays of the Z-boson, as discussed by Duff and
Zeppenfeld~[\ref\zfour{A. Duff and D. Zeppenfeld, Z. Phys. C53 (1992)
529}], but because of their small center-of-mass energy these processes
are only sensitive to scales $\Lambda$ around 100 GeV, which is low
compared to the typical energy scales of scattering processes in hadron
colliders. Therefore, in order to obtain good bounds on the scale of
\gcub\ it is necessary to compute its effect on processes in hadron
colliders at the next order, either in $1/{\Lambda}^2$ or in \alphas.

The former option was pursued by Simmons~[\simmons], and later by
Dreiner {\it et al.}~[\ref\ddz{H. Dreiner, A. Duff and D. Zeppenfeld,
Phys. Lett. B282 (1992) 441}]. Simmons studied the effect of the
gluonic operator on $2\to 2$ parton scattering ($gg\to gg$,
$q\bar{q}\to gg$, $q\bar{q}\to q\bar{q}$ and processes related by
crossing symmetry) at order $1/{\Lambda}^4$. Such contributions come
either from squaring amplitudes with one insertion of \gcub, or from
interfering QCD amplitudes with amplitudes containing two insertions of
\gcub. The main feature of this correction as seen in dijet production,
or alternatively, in inclusive jet production in hadron colliders, is
an excess of events in the high $p_T$ region of the jet
transverse-momentum ($p_T$)
distribution. This signal roughly grows as ${( { {\hat s} /
{\Lambda}^2} )}^2$ where ${\hat s}$ is the parton center-of-mass
energy. It is therefore highly suppressed at low energies, and becomes
appreciable only at energies of the order of \lamb, where unitarity is
violated, and where the parametrization in terms of the effective
operator \gcub\ can no longer be trusted~[\ddz]. Including a
form-factor to unitarize the cross-section reduces the \gcub\ signal at
high energies dramatically, leading to deviations from QCD that are on
the order of the theoretical and experimental uncertainties~[\ddz].

An additional drawback of this approach is that at the order
$1/{\Lambda}^4$ contributions from dimension-8 operators must be
included as well. Since there are many such dimension-8
operators~[\simmons], it is very hard to treat the problem
in full generality.

The difficulties encountered here are related to the fact that the
contribution studied is $O({1/{\Lambda}^4})$. It seems desirable
therefore to investigate the leading-order contribution in
${1/{\Lambda}^2}$, while going to higher order in \alphas. In this case
there are several options to consider. The first is to study the effect
of the gluonic operator on  {\it inclusive} jet production in hadron
colliders at order ${1/{\Lambda}^2}$ and next-to-leading order in
\alphas. Three types of contributions appear at this order. The first
is the interference of $2\to 2$ tree amplitudes containing one
insertion of the gluonic operator, with $2\to 2$ QCD {\it loop}
amplitudes. This contribution is easy to obtain since the relevant QCD
loop helicity amplitudes have been calculated using both string
techniques~[\ref\bk{Z. Bern and D. A. Kosower, Phys. Rev. Lett. 66
(1991) 1669; Nucl. Phys. B379 (1992) 451}] and Feynman
diagrams~[\ref\kunszt{Z. Kunszt, A. Signer and Z. Trocsanyi,
preprint ETH-TH/93-11, hep-ph/9305239}],
and the helicity amplitudes containing one insertion of
the gluonic operator are readily calculable with the use of the
helicity basis. The second contribution involves the interference of
QCD $2\to 2$ tree amplitudes with {\it loop} amplitudes containing one
insertion of the gluonic operator. The latter amplitudes are hard to
compute. Finally, the third contribution arises from the interference
of QCD $2\to 3$ tree amplitudes
($gg\to ggg$, $gg\to q\bar{q}g$, $q\bar{q}\to
q\bar{q}g$ and processes related by crossing symmetry)
with $2\to 3$ tree amplitudes containing one insertion of the gluonic
operator.

In addition to the technical difficulty of calculating the second
contribution, this approach suffers from a more serious problem. A
rough estimate of the order of magnitude of (i) the QCD cross-section,
(ii) the correction calculated by Simmons ($O({1/{\Lambda}^4})$), and
(iii) the $O({1/{\Lambda}^2})$ correction suggested above, reveals that
the $O({1/{\Lambda}^2})$ correction only dominates over the
$O({1/{\Lambda}^4})$ correction at low energies, where they are both
insignificant compared to QCD. This is largely due to a factor
of $\alpha_s/(4\pi)$ suppressing the loop amplitudes.
Instead, in this work we use the third
contribution mentioned above, namely the interference of QCD $2\to 3$
tree amplitudes with $2\to 3$ tree amplitudes containing one insertion
of the gluonic operator, in order
to probe the gluonic operator through {\it
exclusive three jet} production in hadron colliders.
In three-jet production, the \gcub\
correction is a leading order correction, both in ${1/{\Lambda}^2}$ and
in \alphas.

The use of three jet events in hadron colliders to test QCD has
been discussed in the past ~[\ref\greeks {E. Argyres, G. Katsilieris,
C. Papadopoulos, and S. Vlassopulos,
Int. J. Mod. Phys. A7 (1992) 7915},\ref\karliner
{J. Ellis, I. Karliner and W. Stirling, Phys. Lett. B217 (1989) 363}].
Here we propose three different
ways for studying the \gcub\ signal in three
jet events. The first involves the region in which the three jets are
well separated, the second involves the region in which two of the jets
are almost collinear, and the third involves the cross-over between
these two regions. The first approach is probably the easiest one from
the point of view of the experiment but it leads to smaller \gcub\
signals.

The second approach is perhaps the most interesting from a theoretical
point of view. In the region in which two of the partons are almost
collinear, we use a special angular dependence of the \gcub\ signal to
isolate it from QCD. We discuss this approach in detail in section~3
and section~4 but we briefly outline the idea here. When two of the
partons are almost collinear the three-jet event looks almost like a
two-jet event, where one of the partons has a momentum $p$ which is
the sum of the two collinear momenta. The QCD cross-section and the
\gcub\ correction to it
have different properties under azimuthal rotations
of the collinear momenta around $p$, with $p$ held fixed.
An azimuthal dependence of the cross-section in the collinear region
requires a linear polarization of the ``effective gluon''
which replaces the two collinear gluons in the resulting four-parton
cross-section. However, a consequence of the supersymmetric Ward
identities (SWI)~[\ref\swig{M.T. Grisaru, H.N. Pendleton and P. van
Nieuwenhuizen, Phys. Rev. D15 (1977) 996\semi
M.T. Grisaru and H.N. Pendleton,
Nucl. Phys. B124 (1977) 81},\ref\swimp{M.
Mangano and S.J. Parke, Nucl. Phys. B299 (1988) 653},\ref\nonsusy{S.
Parke and T. Taylor, Phys. Lett. B157 (1985) 81}]
is that two of the three independent
four-gluon helicity amplitudes, and one of the two independent
two-quark two-gluon amplitudes vanish for tree-level QCD.
It follows that for fixed helicities of the five external partons
there is only one possible effective gluon helicity; it cannot be
linearly polarized, and so the tree-level QCD cross-section
has no dependence on azimuthal rotations of two collinear partons.
On the other hand, after adding the \gcub-induced amplitude, both
helicities contribute, giving rise to a linear polarization of the
effective gluon and a characteristic azimuthal dependence of the
cross-section.
Precisely the same orthogonality
of the helicity structures of the QCD and the \gcub-induced
four-parton amplitudes, which prevents the appearance of the \gcub\
signal at leading-order, also causes these different behaviors in the
collinear region, thus providing a means for separating the \gcub\
signal from the QCD background.
In measuring the azimuthal dependence of the collinear three-jet
cross-section, one simultaneously probes for the existence of the
\gcub\ operator, and tests the helicity structure of QCD and thereby
the supersymmetric Ward identities.

As is well known, the calculation of QCD amplitudes is not a trivial
task. This task becomes even more daunting with the inclusion of the
gluonic dimension-6 operator, which leads to cumbersome vertices for
three-gluon, four-gluon and five-gluon couplings, all of which would be
needed here. It is therefore crucial to use methods which simplify the
calculation. This work illustrates the power of several such tools,
namely, color ordering~[\ref\color{F. A. Berends and W. T. Giele, Nucl.
Phys.
     B294 (1987) 700\semi
     M. Mangano, S. J. Parke and Z. Xu, Nucl.\ Phys.\ B298 (1988)
     653\semi
     M. Mangano, S. J. Parke, Nucl.\ Phys.\ B299 (1988) 673\semi
     D. Zeppenfeld, Int. J. Mod. Phys. A3 (1988) 2175}]
the helicity basis~[\ref\helicity{
   F.\ A.\ Berends, R.\ Kleiss, P.\ De Causmaecker, R.\ Gastmans,
   and T.\ T.\ Wu, Phys.\ Lett.\ B103 (1981) 124\semi
   P.\ De Causmaeker, R.\ Gastmans,  W.\ Troost,
   and  T.\ T.\ Wu, Nucl. Phys. B206 (1982) 53}%
  \eref{R.\ Kleiss and W.\ J.\ Stirling,
   Nucl.\ Phys.\ B262 (1985) 235\semi
   J.\ F.\ Gunion and Z.\ Kunszt, Phys.\ Lett.\ B161 (1985) 333\semi
   R.\ Gastmans and T.T.\ Wu,
   {\it The Ubiquitous Photon: Helicity Method for QED and QCD}
   (Clarendon Press) (1990)},
\ref\mangano{M. Mangano and S. J. Parke, Phys. Rept. 200 (1991) 301}],
and collinear factorization~[\ref\bergiele{F. A. Berends
and W. T. Giele, Nucl. Phys. B306 (1988) 759},\mangano]
for this extension of QCD.
Thus for example, the orthogonality of QCD amplitudes and amplitudes
induced by the gluonic operator at lowest order is immediately derived
and becomes very transparent  using the helicity basis. Calculating
the new five-gluon vertex becomes unnecessary since we can infer the
five-parton amplitudes from the four-parton amplitudes using collinear
factorization. For some of the calculations we can
extract the contribution of the effective operator by
thinking of it as induced by a heavy colored particle in the loop,
taking advantage of existing string-based QCD loop
calculations~[\bk,\ref\gluloop{Z. Bern, L. Dixon and
D. Kosower, Phys. Rev. Lett. 70 (1993) 2677}].

This paper is organized as follows: in section~2 we explain our notation
and discuss the
calculation of the new amplitudes. We give analytic expressions for the
new four-parton and five-parton amplitudes, and for the corrections to
the cross-section at the parton level. We discuss the behavior of these
corrections in the soft and collinear regions of phase space in
section~3. We then go on to describe the hadron-level calculation, and
suggest different ways of looking for the \gcub\ signal in three jet
events in section~4. We conclude with a summary of the results
in section~5. The appendix reviews
properties of color-ordered amplitudes and collinear factorization,
and shows that \gcub\ does not induce any
collinear singularities on top of the QCD singularities.

\vskip0.2truein
\line{ {\sl 2. Calculating the new amplitudes and interference
terms}\hfill }
\vskip0.1truein

The operator \gcub\  induces new three-, four-, five- and six-gluon
vertices. The three-gluon vertex has the same power of the strong
coupling, $g_s$, as the QCD vertex, and the other new vertices acquire
an additional factor $g_s$ with each additional gluon. All the new
vertices have ${1/{\Lambda}^2}$ multiplying them. To compute the effect
of \gcub\ on three-jet production in hadron-colliders, one needs to
evaluate five-parton tree amplitudes with one insertion of the new
operator, i.e., one new three-, four- or five-gluon vertex. We denote
these new amplitudes by the super-script $(\Lambda)$.

We now briefly describe our notation and the methods used to
organize the calculation: color ordering, the helicity basis and
collinear factorization. In this we follow closely the review by
Mangano and Parke~[\mangano].

Tree amplitudes of $SU(N)$ gauge theories can be organized in a simple
color basis~[\mangano].
Each amplitude is written as a sum of color factors
multiplied by color sub-amplitudes which correspond to specific
color-orderings. Since all the color dependence is factored out of the
sub-amplitudes, and since the color basis is orthogonal at leading
order in the number of colors, each sub-amplitude is separately gauge
invariant and can be calculated using any gauge choice. The $n$-gluon
amplitude ${\cal A}_n$ can be written as,
$$
{\cal A}_n\  =\  \sum_{\{12..n\}^\prime}
{\rm tr}(T^{a_1}T^{a_2}\cdots T^{a_n} )\
A_n(1,2,..,n) \ . \eqn\ngluon
$$
Here
$A_n(1,2,..,n) =  A_n(p_1,\lambda_1;p_2,\lambda_2;\cdots ;p_n,
\lambda_n)$ and $p_i$, $\lambda_i$ are the momentum and
helicity of the $i$-th gluon. The sum in~(\ngluon) is over
non-cyclic permutations of $\{1,2,..,n\}$. We discuss some useful
properties of the color sub-amplitudes in the appendix.

Amplitudes with $n$ gluons and one quark pair can be written as,
$$
{\cal A}_n\ =\ \sum_{\{12..n\}}
{(T^{a_1}T^{a_2}\cdots T^{a_n} )}_{i{\bar {j}}}\
A_n(q;1,2,..,n;{\bar q}) \ .
\eqn\oneqpair
$$
Here $A_n(q;1,2,..,n;{\bar q}) = A_n(q^\lambda;
p_1,\lambda_1;p_2,\lambda_2;\cdots ;p_n, \lambda_n;
{\bar q}^{{\bar\lambda}})$,
$q$, $\lambda$ are the quark momentum and helicity,
${\bar q}$, ${\bar \lambda}$ are the anti-quark momentum and helicity,
$p_i$, $\lambda_i$ are the momentum and helicity of the
$i$-th gluon, and the sum is over all permutations of the $n$ gluons.

We will also need amplitudes with two quark pairs and one
gluon. These we write as,
$$\eqalign{
{\cal A}_5\ =\ &A_5(q_1,{\bar q}_1;q_2,{\bar q}_2;k)\
{1\over N}\delta_{i_1{\bar i}_1} T^a_{i_2{\bar i}_2 }\ +\
 A_5(q_1,{\bar q}_2;q_2,{\bar q}_1;k)\
\delta_{i_1{\bar i}_2} T^a_{i_2{\bar i}_1 }\ +\cr +\
&A_5(q_2,{\bar q}_2;q_1,{\bar q}_1;k)\
{1\over N}\delta_{i_2{\bar i}_2} T^a_{i_1{\bar i}_1 }\ +\
 A_5(q_2,{\bar q}_1;q_1,{\bar q}_2;k)\
\delta_{i_2{\bar i}_1} T^a_{i_1{\bar i}_2 } \ ,
}\eqn\twoqpair
$$
where $q_i$, ${\bar q}_i$ are the momenta of the $i$-th quark and
anti-quark, and $k$ is the gluon momentum.

Calculating the new amplitudes is greatly simplified by using the
helicity basis~[\helicity,\mangano]. Each gluon polarization is
represented in terms of the gluon momentum and another lightlike
momentum, the reference momentum. Choosing the reference momentum
as an appropriate combination of the external momenta
(this choice can be made separately for different
color sub-amplitudes) results in many cancellations in the calculation.
With this representation of the polarization vectors,
helicity amplitudes
of massless external partons are
naturally expressed in terms of spinor products, which are
nothing but square roots of the Lorentz invariants up to
phases. The spinor products are denoted as:
$$\eqalign{
\spb{i}.{j}\ &\equiv\ \langle{i+}\vert j-\rangle\ =\
{\bar{u}}_+(k_i)u_-(k_j)\ ,\cr
\spa{i}.{j}\ &\equiv\ \langle{i-}\vert j+\rangle\ =\
{\bar{u}}_-(k_i)u_+(k_j)\ ,\cr
}\eqn\sproducts
$$
where $\vert i+\rangle$ ($\vert i-\rangle$) is a positive (negative)
helicity spinor with the momentum of parton $i$. We give an
explicit representation of the spinor-products in the appendix.

One of the useful properties of color-ordered helicity amplitudes is
their factorization on collinear poles. When the color adjacent partons
$i$, $i+1$, with momenta $k_i$, $k_{i+1}$ become collinear, the
$n$-parton color sub-amplitude factorizes on the $(n-1)$-parton color
sub-amplitudes where the two collinear partons are replaced by one parton
with momentum $p=k_i+k_{i+1}$ and with positive or negative helicity.
We sometimes refer to this parton as the effective parton. Except for
a subtlety in the terms sub-leading in the number of
colors in the four-quark one-gluon amplitudes (discussed below),
each $n$-parton color structure factorizes on a unique
$(n-1)$-parton color structure.
In the
collinear limit $k_i=zp$, $k_{i+1}=(1-z) p$ and~[\bergiele,\mangano],
$$
A_n(k_1,\ldots,k_i,k_{i+1},\ldots,k_n)\ {\rightarrow}^{i\Vert {i+1}}
\ \sum_{\lambda=\pm}\ {\rm Split}_{-\lambda}(i,{i+1})\ A_{n-1}
(k_1,\ldots,(k_i+k_{i+1})^\lambda,\ldots,k_n)\ .
\eqn\split
$$
The splitting function ${\rm Split}_{-\lambda}(i,j)$
depends on the helicity $\lambda$ of the effective parton,
on the helicities of the two collinear partons and on the
momentum fraction $z$, and is
singular as $s_{i,i+1}=(k_i+k_{i+1})^2\to 0$.
The splitting functions are analogues of
the Altarelli-Parisi coefficients~[\ref\ap{G. Altarelli and
G. Parisi, Nucl. Phys. B126 (1977) 298}]
that are associated with
color-ordered helicity amplitudes instead of cross-sections.
Specifically, the absolute value of the splitting-function,
squared and color-averaged, is
equal to the appropriate polarized Altarelli-Parisi coefficient. We
list the relevant splitting functions in the appendix.

Amplitudes with insertions of \gcub\ satisfy~(\split) as well.
When the two collinear gluons $i$, $j$ are attached to the
new three-gluon
vertex, one can show (see appendix) that
the expression for this vertex goes to zero at least as fast as
$s_{ij}$, cancelling the $1/s_{ij}$ propagator pole.
This is related to the fact that the new vertex
has extra powers of momenta in the
numerator compared to the QCD vertex, to compensate for $1/\Lambda^2$.
Thus the only splitting functions which occur for the five-parton
new amplitudes are the standard splitting functions arising from the
QCD vertices and so,
$$
A_n^{(\Lambda)}(k_1,\ldots,k_i,k_{i+1},\ldots,k_n)\
{\rightarrow}^{i\Vert {i+1}}\ \sum_{\lambda=\pm}\
{\rm Split}_{-\lambda}(i,{i+1})\
A_{n-1}^{(\Lambda)}
(k_1,\ldots,(k_i+k_{i+1})^\lambda,\ldots,k_n)\ .
\eqn\splitnew
$$

Collinear factorization provides a check of $n$-parton amplitudes
when the $(n-1)$-parton amplitudes are known. Moreover,
it is often possible to infer the former from the latter~[\ref
\mhv{Z. Bern, G. Chalmers, L. Dixon and D. A Kosower,
preprint SLAC-PUB-6409, hep-ph/9312333}].
In this work we calculate the relevant \gcub-corrected
four-parton amplitudes and then
guess the form of the five-parton amplitudes by requiring that they
have the correct behavior in all collinear limits. There could be,
however, additional terms in the amplitudes which are finite in all the
collinear limits. We have not been able to find any such terms that
also have the correct dimension and for the five-gluon amplitudes also
satisfy the $U(1)$ decoupling equation (see the appendix). Still, to
verify the results obtained from collinear factorization, we calculated
at least one helicity amplitude for each case (5-gluon, 2-quarks
3-gluons, and 4-quarks 1-gluon) using an alternative method. We find
agreement with the results obtained from collinear
factorization; there are no additional finite terms which
are missed when collinear factorization is used to infer the five-parton
\gcub\ amplitudes.

We begin by listing the relevant four-parton
amplitudes. These are obtained from Feynman diagrams using the helicity
basis, and can be written as one-term expressions.

 For the four helicity amplitudes in
 $gg\to gg$ we have, with $g_s=1$,
$$\eqalign{
  A_4^{(\Lambda)}(1^+,2^+,3^+,4^+)\ &=\
  {3i\over\Lambda^2}
     {2s_{12}s_{23}s_{13}\over\spa1.2\spa2.3\spa3.4\spa4.1}\ ,\cr
  A_4^{(\Lambda)}(1^-,2^+,3^+,4^+)
  \ &=\  {3i\over\Lambda^2}
     {-{\spb2.3}^2{\spb3.4}^2{\spb4.2}^2
     \over \spb1.2\spb2.3\spb3.4\spb4.1}\ ,\cr
  A_4^{(\Lambda)}(1^-,2^-,3^+,4^+)
  \ &=\  0\ ,\cr
  A_4^{(\Lambda)}(1^-,2^+,3^-,4^+)
  \ &=\  0\ .}
\eqn\fourg
$$

The corresponding $gg\to gg$ QCD tree amplitudes are:
$$\eqalign{
  A_4(1^+,2^+,3^+,4^+)\ &=\  0,\cr
  A_4(1^-,2^+,3^+,4^+)\ &=\  0\ ,\cr
  A_4(1^+,2^+,3^-,4^-)\ &=\ A_4(1^+,2^-,3^+,4^-)\ =\ i\
  { {\spa3.4}^4\over
      \spa1.2\spa2.3\spa3.4\spa4.1}\ .\cr
  }
\eqn\fourgqcd
$$
Comparing~(\fourg) and~(\fourgqcd) it is easy to see that
the interference vanishes, and there is no \gcub\
induced correction at leading order. Similarly, for $qg\to qg$,
$$\eqalign{
 A_4^{(\Lambda)}(q^+;1^+,2^+;\bar{q}^-)
 \ &=\ {3i\over \Lambda^2}
    { {\spb1.2}^2{\spb{{q}}.1}^2\spb{\bar{q}}.2\spb2.{q}
    \over \spb{\bar{q}}.{q}\spb{{q}}.1\spb1.2\spb2.{\bar{q}}}\ ,\cr
 A_4^{(\Lambda)}(q^-;1^+,2^+;\bar{q}^+)
 \ &= \ {3i\over \Lambda^2}
    {-{\spb1.2}^2{\spb{\bar{q}}.2}^2\spb{q}.1\spb1.{\bar{q}}
    \over \spb{\bar{q}}.{q}\spb{q}.1\spb1.2\spb2.{\bar{q}}}\ ,\cr
 A_4^{(\Lambda)}(q^+;1^+,2^-;\bar{q}^-)
 \ &\ =0    \ ,\cr
 A_4^{(\Lambda)}(q^-;1^+,2^+;\bar{q}^+)
 \ &= \ 0   \ .\cr}
 \eqn\twoqtwog
$$

The $qg\to qg$ QCD tree amplitudes are:
$$\eqalign{
 A_4(q^+;1^+,2^+;\bar{q}^-) \ &=\ 0\ , \cr
 A_4(q^-;1^+,2^+;\bar{q}^+) \ &=\ 0\ ,\cr
 A_4(q^+;1,2;\bar{q}^-) \ &\ =\ i\
{ -{\spa{\bar{q}}.{i}}^3 \spa{q}.{i}\over
\spa{\bar{q}}.{q}\spa{q}.1\spa{1}.{2}\spa2.{\bar{q}} } \ ,\cr
 A_4(q^-;1,2;\bar{q}^+) \ &= \ i\
{ {\spa{q}.{i}}^3 \spa{\bar{q}}.{i}\over
\spa{\bar{q}}.{q}\spa{q}.1\spa{1}.{2}\spa2.{\bar{q}} } \ ,\cr
}  \eqn\twoqtwogqcd
$$
where $i$ is the negative helicity gluon
and again there is no leading-order interference.

An alternative way to obtain the amplitudes in~(\fourg)
and~(\twoqtwog), is to think of the operator
\gcub\  as being induced by a
heavy colored scalar of mass \Ms\  on the order of $\Lambda$,
circulating in a loop, and to
calculate the $1/{M_s}^2$ term in the resulting loop amplitude.
This is not as cumbersome as it may seem, since all the relevant loop
amplitudes have been calculated by Bern and Kosower~[\bk]
using string techniques for the case of a massless
particle in the loop.
The Feynman parameter polynomials for a massive
scalar are identical to those for a massless scalar, since the
derivative interactions are identical. Only the scalar denominator of
the Feynman integral is changed. In the limit of large \Ms, the only
$O( {1/M_s^2} )$ contributions come from the triangle diagrams,
i.e., diagrams with three legs attached to the loop. The resulting
Feynman parameter integrations are simply integrations over polynomials.
Subtleties of ultraviolet and infrared
divergences and coupling-constant shifts, do not occur in these
loop amplitudes because the corresponding QCD tree amplitudes
vanish.
By calculating
$A_4^{(\Lambda)}(q^+;3^+,4^+;\bar{q}^-)$ in both ways, we
extract the ratio between \Ms\ and \lamb,
$\Lambda^2 = 720\pi M_s^2/\alpha_s$.
We then use this ratio to
check that the two methods agree for the remaining amplitudes.

The zeroes in equations~(\fourgqcd) and~(\twoqtwogqcd) are
manifestations of the supersymmetric Ward identities
(SWI)~[\swig,\swimp]
(unlike the zeroes in equations~(\fourg) and~(\twoqtwog), whose origin
is obscure). According to these identities, the only non-vanishing
four-parton amplitudes are the ones with two positive and two negative
parton helicities. For QCD, which is not supersymmetric, the SWI are
satisfied at
tree-level only~[\nonsusy]. Because of the SWI, as we will see in the
following, the five-parton tree-level QCD
cross-section is invariant under
azimuthal rotations of two collinear partons which leave the sum of the
collinear momenta fixed.

We now turn our attention to the five-parton amplitudes. Since we are
interested in the interference of \gcub-induced amplitudes with QCD
tree amplitudes, we have to calculate only the new amplitudes
corresponding to helicity choices for which the QCD tree amplitudes are
non-zero. These are $(+++--)$ and $(++-+-)$ for the five-gluon
amplitudes, $(\pm;++-;\mp)$, $(\pm;+-+;\mp)$ and $(\pm;-++;\mp)$ for
the two-quark-three-gluon amplitudes, and all possible helicity
structures for the four-quark-one-gluon amplitudes.

To illustrate the use of collinear factorization in deriving these
amplitudes, consider the 2-quark 3-gluon amplitude
$A_5^{{(\Lambda)}}(q^+;1^+,2^+,3^-;\bar{q}^-)$. In the limit when
$q$ and $k_1$ become collinear, this amplitude should factorize on the
4-parton amplitude
$A_4^{{(\Lambda)}}(p^+;2^+,3^-;\bar{q}^-)$ with $p=q+k_1$.
Since the latter amplitude is zero,
$A_5^{{(\Lambda)}}(q^+;1^+,2^+,3^-;\bar{q}^-)$ cannot be singular
in this limit. Similarly, it remains finite as gluons $1$ and $2$
become collinear because ${\rm Split}_+(+,+)$ is zero (see the appendix).
However, in the limit where gluons $2$ and $3$ are
collinear with $k_2=zp$ and $k_3=(1-z)p$,
$$\eqalign{
A_5^{{(\Lambda)}}(q^+;1^+,2^+,3^-;\bar{q}^-) \ &\simeq\
{\rm Split}_-(2^+,3^-)\ \times\
A_4^{{(\Lambda)}}(q^+;1^+,p^+;\bar{q}^-)\cr
&=\ {z^{3/2}\over{(1-z)}^{1/2}\spb2.3 }\ 3i{1\over \Lambda^2}
{ \spb1.{p}\spb1.{q}\spb{p}.{q}\over
  \spb{{\bar q}}.{q} }\ .\cr
}
$$
This is satisfied by the expression
$$
A_5^{{(\Lambda)}}(q^+;1^+,2^+,3^-;\bar{q}^-) \ =\
 3i{1\over \Lambda^2}
{ \spb2.{{\bar q}}\spb1.{q}\spb1.2\spb2.{q}\over
  \spb{{\bar q}}.{q}\spb2.3\spb3.{{\bar q}} }\ ,
$$
which also has the correct behavior when
${\bar q}$ and $k_3$, and ${\bar q}$ and $q$ become collinear, and
which agrees with the result obtained from a Feynman diagram
calculation.

In tree amplitudes with zero or one quark pair, the $n$-parton
color sub-amplitude always factorizes on a unique
$(n-1)$-parton sub-amplitude. The situation is slightly
different for amplitudes with two quark pairs, where the color
ordering is not unique at subleading order in the number of colors.
Thus for example as $q_1$ and
${{\bar q}}_2$ become collinear in the
$O(1/N_c)$ sub-amplitude $A^{{(\Lambda)}}(q_1,{{\bar
q}}_1;q_2,{{\bar q}}_2;k)$ there are contributions from
both orderings of the two gluons (the original gluon and the
one replacing the pair $q_1$, ${{\bar q}}_1$) attached to
the remaining quark line. Thus one finds,
$$
A^{{(\Lambda)}}(q_1,{{\bar q}}_1;q_2,{{\bar q}}_2;k)\ \simeq \
{\rm Split}_+({{\bar q}}_1,q_1) \bigl( A(q_2;k,p;{{\bar q}}_2)\ + \
A(q_2;p,k;{{\bar q}}_2) \bigr)\ .
$$

Our results for the \gcub-induced five-parton amplitudes are as
follows.

For $gg\to ggg$,
$$
  A_5^{(\Lambda)}(i^+,j^+,k^+,l^-,m^-)\ =\ {3i\over \Lambda^2}
  { {\spb{i}.{j}}^2 {\spb{j}.{k}}^2 {\spb{k}.{i}}^2 \over
     \spb1.2\spb2.3\spb3.4\spb4.5\spb5.1 } \ ,
\eqn\massfiveg
$$
where $l^-$ and $m^-$ are not necessarily adjacent. This expression was
inferred from collinear factorization. As a check, we modified
the string-based calculations of
five-gluon loop amplitudes with a massless scalar in
the loop~[\gluloop]
to obtain the $1/M_s^2$ contribution from a scalar of mass $M_s$,
which agreed with equation~(\massfiveg).

The $gg\to ggg$ QCD tree amplitude which interferes with~(\massfiveg)
is:

$$
  A_5(i^+,j^+,k^+,l^-,m^-)\ =\ i
  { {\spa{l}.{m}}^4 \over
     \spa1.2\spa2.3\spa3.4\spa4.5\spa5.1 } \ ,
\eqn\fivegqcd
$$
where, as in~(\massfiveg), $l$ and $m$ are not necessarily adjacent.

For the two-quark three-gluon amplitudes we find,
$$
 A_5^{{(\Lambda)}}(q^+;1,2,3;\bar{q}^-)\ =\
  {3i\over \Lambda^2}
    { -{\spb{i}.{j}}^2{\spb{q}.{i}}^2\spb{j}.{q}\spb{\bar{q}}.{j}
    \over \spb{\bar{q}}.q \spb{q}.1 \spb1.2\spb2.3\spb3.{\bar{q}} }
    \ .
\eqn\qgggbqamp
$$
In this formula $i$ and $j$ are the two positive helicity gluons, and
$i$ is closer to $q$ than $j$ is in the color ordering. The formula
for reversed fermion helicities is
$$
A_5^{{(\Lambda)}}(q^-;1,2,3;\bar{q}^+)\ =\
  {3i\over \Lambda^2}
    { {\spb{i}.{j}}^2{\spb{\bar{q}}.{i}}^2\spb{j}.{\bar{q}}\spb{q}.{j}
    \over \spb{\bar{q}}.q \spb{q}.1 \spb1.2\spb2.3\spb3.{\bar{q}} }\ ,
\eqn\qgggbqampflip
$$
where now $i$ is the gluon closer to $\bar{q}$. These amplitudes were
constructed to have the correct collinear limits. As a check we
calculated $A_5^{{(\Lambda)}}(q^-;1^+,2^+,3^-;\bar{q}^+)$ from
Feynman diagrams.

Notice that the amplitude with reversed quark pair helicities
can be obtained by exchanging the
quark and anti-quark and reversing the ordering of gluons along the
quark line; changing the helicities of the quark-pair amounts to
a charge conjugation operation which reverses the color ordering
associated with the quark line. Thus equation~(\qgggbqampflip)
can be obtained from equation~(\qgggbqamp).

The required QCD amplitudes are:
$$\eqalign{
 A_5(q^+;1,2,3;\bar{q}^-)\ &=\
    i{ -\spa{q}.{i}{\spa{\bar{q}}.{i}}^3
    \over \spa{\bar{q}}.q \spa{q}.1 \spa1.2\spa2.3\spa3.{\bar{q}} }
    \ ,\cr
 A_5(q^-;1,2,3;\bar{q}^+)\ &=\
    i{ \spa{\bar{q}}.{i}{\spa{q}.{i}}^3
    \over \spa{\bar{q}}.q \spa{q}.1 \spa1.2\spa2.3\spa3.{\bar{q}} }
    \ ,\cr
  }  \eqn\threegqcd
$$
where $i$ is the negative helicity gluon.

Finally, for the four-quark-one-gluon amplitudes we have,
$$
\eqalign{
 A^{{(\Lambda)}}(q_1^+,\bar{q}_2^-;q_2^+,\bar{q}_1^-;k^+)
 \ &=\  {3i\over \Lambda^2}
    { \spb{q}_1.{{q}_2}\spb{q}_1.{k}\spb{q}_2.{k}
    \over \spb{\bar{q}}_1.{{q}_1}\spb{\bar{q}}_2.{{q}_2}  }\ ,\cr
 A^{{(\Lambda)}}(q_1,\bar{q}_1;q_2,\bar{q}_2;k)
 \ &=\ 0  \ . \cr}
\eqn\qqqqgamp
$$
These amplitudes were inferred from collinear limits,
and verified by a Feynman-diagram calculation. Again, amplitudes
with reversed quark pair helicities can be obtained by exchanging the
quark and anti-quark, and this can be done for each quark-pair
separately.

The corresponding QCD amplitudes are:
$$
\eqalign{
 {\cal A}(q_1^+,\bar{q}_2^-;q_2^+,\bar{q}_1^-;k^+)
 \ &=\ F\
{ \spa{q_2}.{\bar{q_1}}\over
\spa{q_2}.{k}\spa{\bar{q_1}}.{k} }     \ ,\cr
 A(q_1^+,\bar{q}_1^-;q_2^+,\bar{q}_2^-;k^+)
 \ &=\ F\
{ \spa{\bar{q_2}}.{q_2}\over
\spa{q_2}.{k}\spa{\bar{q_2}}.{k} }     \ ,\cr
 } \eqn\onegqcd
$$
where
$$
F\ =\ i\ {{\spa{\bar{q_1}}.{\bar{q_2}}}^2\over
\spa{\bar{q_1}}.{q_1} \spa{\bar{q_2}}.{q_2} }\ .
$$
Amplitudes with reversed quark-pair helicities can be
obtained from~(\onegqcd) by exchanging the appropriate
quark and anti-quark in $F$.

Notice that the expressions in~(\qqqqgamp), (\onegqcd)
were derived for different
quark pair flavors. The amplitudes with same flavor quarks can be
obtained from them by antisymmetrizing over identical quarks.

For all the different parton processes, one can obtain the amplitudes
with all helicities reversed by parity, which exchanges
angle-brackets $\langle\ \rangle$ and square-brackets $[\ ]$
in~(\massfiveg)\ --~(\qqqqgamp), and multiplying by an
additional minus sign for amplitudes with an odd number of gluons.

We now need to interfere these amplitudes with the appropriate QCD
amplitudes, and to sum over colors and helicities. Since the
initial and final states vary from process to process,
we perform the initial state averaging later.

For $gg\to ggg$ we get, summing over final states colors and helicities,
$$\eqalign{
\delta \sigma^{(\Lambda)}_{ggggg}\ &\equiv\
\sum_{\rm helicities}\sum_{\rm colors}
\left(A_5^*A_5^{{(\Lambda)}}+A_5
{A_5^{{(\Lambda)}}}^*\right)
\ =\cr
&=\ -6g_s^6 {1\over \Lambda^2} N_c^3 (N_c^2-1)
\ \times\cr \times\
\bigg[ \sum_{\{2,3,4,5\}}
&{\left(\spb1.2\spb2.3\spb3.4\spb4.5\spb5.1\right)}^{-2}
{\sum _{l,m\in\{1,2..,5\}} } _{{l\neq m}}
{ \left( \spb{i}.{j} \spb{j}.{k} \spb{k}.{i}
{\spb{l}.{m}}^2 \right) }^2\ +c.c.\bigg]\ , }
\eqn\hcsfivg
$$
where the first summation is over all permutations of $(2,3,4,5)$.

For scattering processes involving four quarks and one gluon,
the interference term summed over colors and helicities is,
$$\eqalign{
\delta\sigma^{(\Lambda)}_{q\bar{q}q\bar{q}g}\ &=\
6 g_s^6 {1\over \Lambda^2} N_c (N_c^2-1)
\ {1\over {\spa{{{\bar q}_1}}.{{q_1}}^2
\spa{{{\bar q}_2}}.{{q_2}} }^2}\
\left(
{ \spa{{{\bar q}_1}}.{{q_2}}\over
 \spa{{{\bar q}_1}}.{k} \spa{{q_2}}.{k} }-
{ \spa{{{\bar q}_2}}.{{q_1}}\over
 \spa{{{\bar q}_2}}.{k} \spa{{q_1}}.{k} } \right) \ \times\cr
&\times \
\bigg[\
{ \spa{{{\bar q}_1}}.{{{\bar q}_2}} }^2 \spa{{q_1}}.{{q_2}}
\spa{{q_1}}.{k} \spa{{q_2}}.{k}\ +\
{ \spa{{q_1}}.{{{\bar q}_2}} }^2 \spa{{{\bar q}_1}}.{{q_2}}
\spa{{{\bar q}_1}}.{k} \spa{{q_2}}.{k}\ +\cr &\ \ \ +
{ \spa{{{\bar q}_1}}.{{q_2}} }^2 \spa{{q_1}}.{{{\bar q}_2}}
\spa{{q_1}}.{k} \spa{{{\bar q}_2}}.{k}\ +\
{ \spa{{q_1}}.{{q_2}} }^2 \spa{{{\bar q}_1}}.{{{\bar q}_2}}
\spa{{{\bar q}_1}}.{k} \spa{{{\bar q}_2}}.{k}
\ +\ c.c. \bigg] \ .   \cr }
\eqn\hcsoneg
$$
Again, equation~(\hcsoneg) holds for different quark pair flavors,
but the modification for same quark flavors is straightforward.

The corresponding expressions for scattering processes involving two
quarks and three gluons are more complicated, so we only give here the
analytic expression for the color sum, without summing over helicities.
(We perform the latter summation numerically.)
$$\eqalign{
\delta\sigma^{(\Lambda)}_{q\bar{q}ggg}\ &=\
{ (N_c^2-1)\over N_c^2 } \sum_{\rm helicities}
\bigg[\ {(N_c^2)}^2 \sum_{\{1,2,3\}}
A^{{(\Lambda)}}(1,2,3) A^*(1,2,3) \ + \cr
&+ \ N_c^2 \sum_{\{1,2,3\}}
A^{{(\Lambda)}}(1,2,3)\
\Big(-2A^*(1,2,3)-A^*(2,1,3)-A^*(1,3,2)\ +\cr
&+A^*(3,2,1) \Big)\
+ \ \Big(\sum_{\{1,2,3\}}
A^{{(\Lambda)}}(1,2,3) \Big)
 \Big(\sum_{\{1,2,3\}}
 A^*(1,2,3) \Big) \ +c.c.\bigg]\ .  \cr }
\eqn\csthreeg
$$
Here all summations are over the six permutations of $(1,2,3)$, and
$A(1,2,3)$ is shorthand for $A(q;1,2,3;{\bar q})$.


\vskip0.2truein
\line{ {\sl 3. Collinear and soft behavior of the \gcub\ correction to
the cross-section}\hfill }
\vskip0.1truein

As is well known, the leading order QCD cross-section for three jet
production is singular in the soft and collinear regions of
phase-space. This property is exhibited as strong peaking of the QCD
distributions near values of the three-jet
event variables which correspond
to a collinear or to a soft region. In order to search for the
\gcub\ signal in three jet events
it is important to understand its behavior in these regions.
As mentioned in the introduction, the special angular dependence of the
\gcub\ signal in collinear regions
can in principle be used to separate it from QCD. In this section we
will explain this angular dependence in more detail. First, however, we
discuss the behavior of the \gcub\ signal in soft regions.

Inspecting the expressions for the interference of QCD amplitudes and
\gcub-induced amplitudes reveals that the different parton energies
appear with the same power in the numerator and the denominator in each
term, so that the interference remains finite as any of the energies
approaches zero. Thus the \gcub\  correction to the three
jet cross-section
remains finite in soft regions of phase-space. Appropriate cuts should
therefore be imposed to avoid these regions in searches for \gcub\
since the singular QCD background will swamp the signal there.

As for collinear regions, we know that the new amplitudes have
collinear singularities
since these were used to construct them. For both the QCD and the new
amplitudes, the singularity associated with partons $i,j$ becoming
collinear is of the form ${\spa{i}.{j}}^{-1}$ or ${\spb{i}.{j}}^{-1}$,
so that its magnitude is ${s_{ij}}^{-{1\over2}}$. Therefore it results
in an ${s_{ij}}^{-1}$ singularity in the QCD cross-section.

In the following collinear analysis, we hold fixed the helicities
of the five external partons, and examine the possible helicities
of the effective parton with momentum $k_i+k_j$.

If only one of ${\rm Split}_+(i,j)$ and ${\rm Split}_-(i,j)$ is
non-zero, the resulting singularity in the \gcub-corrected
cross-section is only ${s_{ij}}^{-{1\over2}}$, because the helicity
structures of the four-parton QCD amplitudes and the new amplitudes are
orthogonal, so that only one of the two amplitudes in the interference
is singular.
However, for $g g \to g$ and $q\bar{q}\to g$
both ${\rm Split}_+(i,j)$ and ${\rm
Split}_-(i,j)$ can be non-zero, in which case
the \gcub\ correction to the
cross-section has a single pole in $s_{ij}$. Then, the QCD
cross-section and the new cross-section are equally singular in the
limit when two partons become collinear. Still, the form of the
singularity is different. The singular term in the tree level
five-parton QCD
cross-section is always of the form ${\vert{\rm Split}_\pm(i,j)\vert}^2$
because there is only one choice of the effective parton helicity that
yields a non-zero four-parton amplitude, due to the SWI zeroes in
equations~(\fourgqcd) and~(\twoqtwogqcd).
In contrast, in the interference of a QCD
amplitude with a \gcub\  amplitude, the choice of the effective parton
helicity is still unique in each amplitude,
but it must be of the opposite helicity for the QCD
amplitude as for the \gcub\  amplitude, due again to four-parton
orthogonality. Therefore, the singularity is always
of the form ${{\rm Split}_\pm(i,j)}^* {{\rm Split}_\mp(i,j)}$. This
last factor carries the phase $2\varphi$, where $\varphi$ is the
angle associated with azimuthal rotations of the two collinear momenta
around the direction of their sum, $p$, with $p$ held fixed.
Notice that the phase has to be $2\varphi$ rather than $\varphi$ or it
would matter which of the two partons $i$, $j$ we use to determine
$\varphi$ (exchanging $i$ and $j$ amounts to exchanging $\varphi$ and
$\pi-\varphi$). This is the only $\varphi$ dependence of the
cross-section in the collinear region, as the rest of the expression is
just the product of the QCD and \gcub\  four-parton amplitudes and they
only depend on the four-parton parameters.

As we mentioned above, the trivial $\varphi$-dependence of the QCD
cross-section follows from the supersymmetric Ward identities. The QCD
cross-section has no $\varphi$ dependence because of the zeroes in
equations~(\fourgqcd) and~(\twoqtwogqcd), and these, in turn, are
a consequence of the SWI.

The $\varphi$ dependence of the cross-section is related to the
polarization state of the effective parton.
Each QCD five-parton amplitude factorizes on just one four-parton
amplitude with either a positive or a negative {\it circularly}-polarized
parton. The resulting QCD cross-section can therefore have no $\varphi$
dependence. In contrast, each \gcub-corrected five-parton amplitude
factorizes on the sum of two four-parton amplitudes with
oppositely polarized gluons. The resulting cross-section contains an
interference term corresponding to a {\it linearly}-polarized effective
gluon, and this term gives rise to a non-trivial $\varphi$ dependence,
correlated with the polarization vector.

Since the singular part of the \gcub\ correction behaves as
$e^{2i\varphi}$ it is washed out upon integrating $\varphi$ between zero
and $2\pi$. The \gcub\ correction to the total cross-section is
therefore finite; it has neither soft nor collinear
singularities \footnote\dag{This implies that the four-parton loop
amplitudes with one insertion of the effective operator,
which were mentioned in the introduction, are finite.
There are neither ultraviolet nor infrared divergences
(soft nor collinear) for them
to cancel against. In addition, these loop diagrams have no imaginary
parts, as can be seen by cutting them into two tree diagrams and
using the orthogonal helicity
structures of the QCD and \gcub\ four-parton tree diagrams.
So they are apt to be quite simple expressions.}.

The non-trivial $\varphi$ dependence of the \gcub\ signal
can be used to separate it from tree level QCD.
If we consider three jet events in a region in which two of the partons
are almost-collinear, it is always possible
to choose a frame in which the $\varphi$ dependence of the \gcub\
correction is $\cos 2\varphi$.
If we then weight the events by $\cos 2\varphi$ and
integrate over $\varphi$, tree level QCD washes out while the \gcub\
contribution remains non-zero. It is useful to define the $\cos 2\varphi$
expectation value,
$$
\langle\cos 2\varphi\rangle\equiv\
{ \int\sum \left(M(M^\prime)^*\cos 2\varphi\ +\ c.c.\right) \over
\int\sum \left(M(M^\prime)^*\ +\ c.c.\right) }\ ,
\eqn\vev
$$
where $M$ and $M^\prime$ are either the QCD or the \gcub-induced
five-parton amplitudes (depending on whether one is considering the
pure QCD signal or the \gcub\ correction to the signal), the sum is over
colors, helicities and the different relevant parton processes, and the
integration is restricted to the almost-collinear region of phase space
and sweeps the full range of $\varphi$, $0\leq\varphi\leq 2\pi$.
This expectation value is a good probe of the operator \gcub; it
receives contributions from the \gcub\ correction and no
contribution from leading-order QCD. Furthermore, it provides a test of
the SWI, or alternatively, of the helicity structure of tree-level QCD.

Notice that four-quark contact operators do not contribute to
$\langle\cos2\varphi\rangle$ at leading-order.
These operators only give rise to
four-quark one-gluon amplitudes, which can only factorize on four-quark
amplitudes containing the four-quark contact
vertex, when a quark and a gluon
become collinear. Since the helicity of the effective
quark is determined by the helicity of the original quark, the same
splitting function occurs in both these new amplitudes and the
corresponding
QCD amplitudes so that the interference has no $\varphi$ dependence.
Thus, these operators do not contribute to the $\cos 2 \varphi$
expectation value.

However, some background to the \gcub\ signal as
measured through this expectation value will arise from
higher order QCD effects. A similar azimuthal dependence
to the one described above occurs in next-to-leading-order corrections
to $2\to 3$ scattering and in
tree-level $2\to 4$ scattering in pure QCD. The first effect turns out
to be very small.
The collinear behavior of five parton loop amplitudes is similar
to equation~(\split), except there are now two terms~[\ref
\loopsplit{Z. Bern, L. Dixon, D. Dunbar and D. A. Kosower,
in preparation}]. In the first term a ``loop splitting-function''
multiplies four-point tree amplitudes; here the previous tree-level
arguments still forbid an azimuthal dependence. In the second term, the
usual tree splitting-function multiplies a four-parton QCD loop
amplitude; the only azimuthal dependence arises when the loop amplitude
is one which vanishes at tree-level.
The relative magnitude of this effect compared to the \gcub\  signal
is given by the relative magnitudes of the \gcub\ correction to
the four-parton amplitude and the relevant (infrared
and ultraviolet finite) four-parton QCD one-loop amplitude.
The relevant four-gluon QCD loop amplitude, for example, is~[\gluloop]
$$
A_4^{1-{\rm loop}}(1^-,2^+,3^+,4^+)\ =\ {i\over 48\pi^2}
N_c\left(1-{n_f\over N_c}\right)
{ \spa2.4{\spb2.4}^3
\over \spb1.2\spa2.3\spa3.4\spb4.1 }\ .
$$
Including the coupling constants,
it is smaller than the \gcub\ amplitude~(\fourg) by a factor of
the order of $(\Lambda^2/{\hat s})(\alpha_s/ 18\pi)$.
For the values of $\Lambda$
and ${\hat s}$ relevant for the present analysis (see section 4),
this ratio is of the order of
$10^{-3}-10^{-2}$. This estimate is essentially the same as the one
mentioned in the introduction where it was remarked that
the interference of a QCD loop
amplitude with a \gcub\ amplitude is small compared with the square of a
\gcub\ amplitude. A similar estimate holds for quark-gluon amplitudes.

The second source of background, $2\to 4$ scattering in pure QCD, is
harder to estimate. In this case too there could be a non-trivial
dependence on azimuthal rotations of two almost collinear partons,
but to get a large correction to the three-jet cross-section
one of the remaining two final state partons should be soft or collinear.
However, in this limit the six-parton amplitude factorizes on the
five-point tree amplitude so there is again
a unique possibility for the helicity of the effective parton
leading to the conclusion that $2\to 4$ QCD contributions to
$\langle\cos 2\varphi\rangle$ are probably small.


\vskip0.2truein
\line{ {\sl 4. The \gcub\ signal in three jet events}\hfill }
\vskip0.1truein

We can now proceed to calculate the effect of \gcub\ on different
distributions in three jet variables. We specifically consider
here three jet production at the Tevatron as an example.
Since we
are only interested in a qualitative estimate of the \gcub\ signal, we
take the jet momenta to coincide with the parton momenta, without using
any hadronization algorithm. We still have to fold the partonic
cross-section with the proton, or anti-proton, structure functions:
$$
d\sigma \ =\ \sum_i \int dx_1dx_2 F_1(x_1,Q^2) F_2(x_2,Q^2)
d{\hat\sigma}_i\ ,
\eqn\hadron
$$
where $i$ labels the different partonic processes,
and we choose the scale $Q$ as $Q^2 = (x_1x_2 s)$. We use the
DFLM~[\ref\dflm{M. Diemoz, F. Ferroni, E. Longo and G. Martinelli, Z.
Phys. C39 (1988) 21 (set 1)}]
structure functions for the calculation.
Here $d{\hat\sigma}_i$ stands for either the QCD cross-section or the
\gcub\ correction. We evaluate the two simultaneously, which allows for
some checks on the calculation by comparing our results for pure QCD
to existing simulations of the QCD distributions (we used the CDF
collaboration's results as reported in~[\ref\cdf{CDF collaboration, F.
Abe {\it et al.}, Phys. Rev. D45 (1992) 1448}]). Since there is no mass
scale in the problem besides $\Lambda$
(we neglect quark masses), both the QCD
cross-section and the \gcub\ correction have definite energy scaling
properties; the QCD cross-section is energy-independent, and the
correction scales as the energy squared. This can be used to factorize
the calculation into two parts: an angular integration over final
states, and an integration over the incident partons' $x_1$ and $x_2$.
This leads to a significant reduction in computer time.
We perform the first integral using the Monte-Carlo program
SAGE~[\ref\sage{J. H. Friedman, J. Comp. Physics, 7 (1971) 1}].

We also take into account the renormalization of \gcub\ between the
scales $\Lambda$, where \gcub\ appears with a coefficient of $4\pi$
(see equation~(\efflag)), and the scale $Q$~[\simcho].
The anomalous dimension of \gcub\ is equal to
$\lambda g_s^2/ 8\pi^2$ with
$\lambda\ =\ 7+2n_f/3$~[\ref\morozov{A. Y. Morozov,
Sov. J. Nucl. Phys. 40 (1984) 505},\ref\narison{S. Narison and
R. Tarrach, Phys. Lett. B125 (1983) 217}]
so that the coefficient of \gcub\
at the scale $Q$ is,
$$
C(Q)\ =\ 4\pi {\left({\alpha_s(Q)\over\alpha_s(\Lambda)}\right)}^
{{\lambda\over 2b}}\ ,
\eqn\coeffic
$$
where $b\ =\ -11/2+n_f/3$ and,
$$
\alpha_s(Q)\ =\ {\alpha_s(M_z)\over
1-b{\alpha_s(M_z)\over\pi}\log{Q\over M_z} }\ ,
$$
where we take $\alpha_s(M_z)=0.118$.
The resulting
reduction in the coefficient of \gcub\ for the relevant values of
$\Lambda$ and $Q$ (see the following) is roughly ten percent.
The operator \gcub\ does not mix with any other
dimension-6 operator through the renormalization group
equation~[\morozov,\narison]. In this sense, studying it separately from
the other operators is not inconsistent.

We now discuss the geometry of three-jet events (see fig.~1),
the cuts we impose
and the different distributions we study. We use the notation of the
CDF collaboration~[\cdf].
Viewed in the center-of-mass system,
the three outgoing jets can be described by five
independent variables; the
total energy of the jets $\sqrt{{\hat s}}$, and four scale-independent
quantities, namely, the energy fractions of two of the jets, and two
angles. Labeling the jets 3, 4 and 5, in order of decreasing energies,
the energy fraction of jet $i$ is defined as
$$
x_i={2E_i\over \sqrt{{\hat s}}}\ ,\ \ \ \ \ \ \ \ \ \ i=3,\ 4,\ 5 \ ,
\eqn\fractions
$$
where $E_i$ is the energy of the $i$-th jet, and $x_3+x_4+x_5=2$. The
angles we use are the angle between the fastest jet and the beam
direction, $\theta$, and the angle between the jets' plane and the plane
of the fastest jet and the beam, $\psi$. An additional angle, the
overall azimuthal angle around the beam direction, is required for a
full description of
the event, but the dependence on it is trivial for unpolarized beams.

We describe here three different ways of probing \gcub. The first is
to study distributions in various three jet variables in the region
where the three jets are hard and well separated. The second is to study
the $\cos 2\varphi$ expectation value of equation~(\vev)
obtained by weighting the relevant matrix elements by
$\cos2\varphi$ in the region where jets 4 and 5
are almost collinear as discussed in section~3.
Finally, one can interpolate between these two
regions and study distributions which reflect the qualitative
differences between the QCD cross-section and the \gcub\ correction in
the cross-over region.

To ensure that the three jets are well separated we impose the following
set of cuts (set A)
$$
\eqalign{
\sqrt {{\hat s}}\ &\geq\ 250\ {\rm GeV}\ ,\cr
x_3\ &\leq\ 0.8\ ,\cr
\vert\cos\theta\vert\ &\leq\ 0.8\ ,\cr
30^\circ\ &\leq\ \psi\ \leq\ 150^\circ\ .\cr}
\eqn\seta
$$
We take the lower bound on the total jet energy,
$\sqrt {{\hat s}}$, to be 250 GeV for the Tevatron
following ref.~[\cdf].
Increasing this bound enhances the deviation from
QCD, but decreases the cross-section, thus leading to poorer
statistics.
The energy fraction of the fastest
jet, $x_3$, varies between ${2\over 3}$ and $1$. The former value
corresponds to a symmetric event, and the latter to either a soft jet
(5) or to two collinear jets (4 and 5). It is therefore necessary to set
some cut on the maximum value of $x_3$ to avoid soft or collinear jets.
As this cut decreases, the sensitivity of the signal to the
deviation from QCD improves, since it drives the events
away from the soft
region, where QCD is divergent whereas the deviation is finite. We take
the $x_3$ cut to be 0.8. To avoid regions where jet 3 becomes
collinear with the beam direction, we require that
the absolute value of $\cos\theta$ be smaller than 0.8.
Finally, to prevent jets 4 and 5 from approaching
the beam direction we take $\psi$ to lie between $30^\circ$
and $150^\circ$.

The distributions in $x_3$, $\psi$ and the three
jet invariant mass $\sqrt{{\hat s}}$ for the scale
$\Lambda=1\ {\rm TeV}$ are plotted in Fig. 2. The errors
bars shown are the estimated statistical errors assuming an
integrated luminosity of $15 {\rm pb}^{-1}$.
The \gcub\ correction to all three distributions is positive
throughout this range. The deviations from QCD change sign
because we plot normalized distributions following the CDF conventions.
Even though the shape of
the \gcub\ correction is different from the shape of the QCD
distributions for these variables, when added to the
larger QCD cross-sections they still result in distributions that are
qualitatively similar to the QCD distributions, apart, roughly, from a
vertical shift. Thus, the effect of \gcub\ on the normalized
distributions is quite small in this region, and although it is
larger than the estimated statistical errors, it seems unlikely that
studying this region by itself could conclusively determine the
presence of the effective operator.

We have also computed distributions in
${\cos}^{-1}({(x_4-x_5)/x_3})$, as suggested
by Ellis, Karliner and Stirling~[\karliner], in
$(1+\cos\theta) /( 1-\cos\theta)$, following
ref.~[\greeks]
and in $x_4$ and $\theta$. None of them improves the sensitivity to the
signal.

The situation becomes even worse if one includes a form-factor in the
analysis. The operator \gcub\ violates unitarity in $2\to 2$ parton
scattering as discussed by Dreiner {\it et al.}~[\ddz].
These authors replace
$1/\Lambda^2$ by
$$
{1\over \Lambda^2} {1\over {(1+4\pi {\hat{s}}/(3\Lambda^2))}^2 }
\eqn\formf
$$
where ${\hat{s}}$ is the parton center-of-mass energy,
to restore unitarity in most partial
waves in the $2 \to 2$ scattering \footnote\dag{Here we modified the
numerical factor to take into account the different scale definitions.
The scale $\Lambda$ used by Dreiner {\it et al.} is equal to
our $\Lambda/\sqrt{16\pi}$.}.
Since the \gcub\ signal in dijet
production is $O({\hat{s}}^2/\Lambda^4)$
it is greatly reduced by the form-factor. In
the case at hand, the effect of the form-factor is smaller
but it still reduces the already-small \gcub\ signal of Fig.~2.

Alternatively, the effective operator can be probed in the second way
mentioned above, namely, by focusing on the collinear region and using
the $\varphi$ dependence of the correction induced by the effective
operator to detect it.
To get to the region where jets 4 and 5 are almost collinear
and neither is soft we take
$x_3\geq 0.95$ and $x_5\geq 0.3$. We need some further cut to ensure
that the two jets are still distinguishable.
If we were to use a standard jet
algorithm which involves
$R=\sqrt{{(\Delta\eta)}^2+{(\Delta\phi)}^2}$,
where $\eta$ is the jet's rapidity and $\phi$ is the usual azimuthal
angle of the jet, in order to acheive this, we would
introduce a spurious $\varphi$-dependence which would spoil the effect
we describe here. Instead it would probably be best to impose a cut on
the minimum transverse momentum of jets 4 and 5 with respect to the
direction of jet 3
\footnote\ddag{We thank J. D. Bjorken for
suggesting this.}
(this direction is the direction of the boost to the
rest frame of jets 4 and 5).
However, since we factorize the calculation into independent
integrations over $x_1$, $x_2$ and over the final states,
we impose the stricter cut:
$$ x_4\sin\theta_{34}\ \geq\ {5 {\rm GeV}\over {(\sqrt{{\hat
s}})}_{min}/2}\ ,
\eqn\strict
$$
where $\theta_{34}$ is the angle between jets 3
and 4. Combined with the cut on $\hat s$, $\hat s\geq {{\hat s}}_{min}$,
this cut guarantees that the transverse momentum of jets 4 and
5 is greater than 5 GeV.

In addition, we take $\vert\cos\theta\vert\leq 0.5$ so that the
2-jet-like event is almost perpendicular to the beam direction and the
\gcub\ correction is more pronounced.
However, we impose no cuts
on the angle $\psi$. This angle should be allowed to sweep its entire
range so that the azimuthal angle $\varphi$
of jets 4 and 5 can vary between zero
and $2\pi$.

We collect these cuts (set B) here
$$
\eqalign{
\sqrt {{\hat s}}\ &\geq\ 250\ {\rm GeV}\ ,\cr
x_3\ &\geq\ 0.95\ ,\cr
x_5\ &\geq\ 0.3\ ,\cr
x_4\sin\theta_{34}\ &\geq\ {5 {\rm GeV}\over {(\sqrt{{\hat
s}})}_{min}/2}\ ,\cr
\vert\cos\theta\vert\ &\leq 0.5\ .\cr}
\eqn\setb
$$

The resulting QCD and \gcub-corrected distributions in $x_3$ and in
the cosine of the angle between jets 4 and 5, $\cos\theta_{45}$, are
plotted in Fig. 3. Again, the error bars shown are the estimated
statistical errors assuming an integrated luminosity of
$15 {\rm pb}^{-1}$.
In 3a and 3b we plot
the distributions obtained by weighting the
matrix elements with $\cos2\varphi$,
$$
{d\sigma_\varphi\over d\cos\theta_{45}}\ =\
\int d\varphi\ \cos 2\varphi\
{d^2\sigma\over d\varphi d\cos\theta_{45}}\ =\
\langle \cos 2\varphi\rangle\
{d\sigma\over d\cos\theta_{45}}\ ,
\eqn\weighta
$$
and,
$$
{d\sigma_\varphi\over dx_3}\ =\
\int d\varphi\ \cos 2\varphi\
{d^2\sigma\over d\varphi dx_3}\ =\
\langle \cos 2\varphi\rangle \
{d\sigma\over dx_3}\ ,
\eqn\weightb
$$
where $\langle\cos 2\varphi\rangle$
of~(\weighta) is different than that of~(\weightb) as
they are distributed in different variables.
In this case the angle $\varphi$ is the
azimuthal angle of jet 4 around the sum of the momenta of jets 4 and 5,
and it is related to the angle $\psi$: $\varphi = \psi-\pi/2$.
The resulting distributions are negative as they receive larger
contributions from events that are co-planar with the beam. The
magnitude of the QCD
distributions decreases on approaching the collinear limit while the
magnitude of the \gcub\ correction grows. In 3c and 3d we show
$\langle\cos 2\varphi\rangle$
as obtained from QCD with and without \gcub. Clearly,
the signal is dominated by the \gcub\ correction. Also shown are the
\gcub-corrected
distributions with the form-factor~(\formf) included. The
form-factor reduces the \gcub\ correction uniformly over the entire
$x_3$ and $\cos\theta_{45}$ ranges. The reduction factor is equal to
the scale-averaged form-factor, and is about one half, which is roughly
the value of the form-factor near the cut
$\sqrt {{\hat s}}\ =\ 250\ {\rm GeV}$, from
where most of the cross-section comes.

Recall that  the \gcub\ correction to the cross-section has
a non-trivial $\varphi$-dependence only if the the collinear
partons are a quark and an anti-quark or two gluons with
opposite helicities. Therefore if quark and gluon jets could be
separated experimentally (a difficult task)
one could get a larger $\langle\cos 2\varphi\rangle$
signal by only considering collinear gluon pairs or
quark anti-quark pairs.

This method provides a distinctive probe of \gcub. The $\cos2\varphi$
expectation value is dominated
by the \gcub\ correction and does not receive
contributions from the tree-level five-parton QCD cross-section or from
dimension-6 operators that can be related to four-quark contact
operators.
One source of background to this measurement is higher-order
contributions to the QCD cross-section which have a non-trivial
dependence on azimuthal rotations in the collinear region. As discussed
earlier, these contributions are probably quite small.
Some azimuthal dependence could also arise from color-strings associated
with the hadronization of the out-going partons. The magnitude of these
effects can be estimated by combining the exact five-parton
cross-section with Monte-Carlo simulations for parton showers and
hadronization, but we have not done so here.
Also, the different dependence of the effective operator contribution
on the parton center-of-mass energy
could be used to separate its signal from these QCD backgrounds. In
addition to these theoretical issues, there are experimental
difficulties associated with measuring the azimuthal dependence of
almost collinear jets. In particular, both the algorithm used to define
these jets and the detector should not induce a significant spurious
azimuthal dependence; the $\cos2\varphi$ asymmetry is only a
few percent for the value of $\Lambda$ considered here.

So far we have considered the effects of \gcub\ in two regions: one
where the three jets are well separated, and the other where two jets
are almost collinear. The \gcub\ correction is
positive in the first region and negative in the second. (The sign of
the effect is reversed if the coefficient of the operator
is negative.) One can use
this fact to enhance the \gcub\ signal by studying {\it normalized}
distributions in a variable which parametrizes the extent of
collinearity, such as the angle between jets 4 and 5 or $x_3$, over a
wide region which includes both the region where the jets are well
separated and the region where they become collinear. Since the
\gcub\ corrections change sign over this region the resulting
normalized distributions are more sensitive to the \gcub\
signal compared to studies which avoid the collinear region altogether.
This is indeed the case as one can see in Fig. 4 where
we plot the normalized distributions in $x_3$
with $x_3$ varying from its minimum value, 2/3, all the way up to
0.95.
Here we combine the cuts on $\cos\theta$ and on $\psi$ of set A~(\seta)
which ensure that the five partons are well separated outside the
collinear region, with the cuts on $x_5$ and on the $p_T$ of jets 4 and
5 of set B~(\setb) so that jets 4 and 5 are still
distinguishable when the collinear region is approached.
We label these cuts as set C,
$$
\eqalign{
\sqrt {{\hat s}}\ &\geq\ 250\ {\rm GeV}\ ,\cr
x_3\ &\leq\ 0.95\ ,\cr
x_5\ &\geq\ 0.3\ ,\cr
x_4\sin\theta_{34}\ &\geq\ {5 {\rm GeV}\over {(\sqrt{{\hat
s}})}_{min}/2}\ ,\cr
\vert\cos\theta\vert\ &\leq\ 0.8\ ,\cr
30^\circ \ &\leq\ \psi\ \leq\ 150^\circ \ .\cr}
\eqn\setc
$$

The deviation due to \gcub\ is more pronounced here than in Fig.~2,
where the $x_3$ distribution is evaluated in the region where the jets
are well separated.
Again, the errors bars shown are the estimated statistical errors
assuming an integrated luminosity of 15 ~${{\rm pb}}^{-1}$.
Except in the cross-over region where it changes sign,
the deviation from QCD is significantly larger than these errors.
For $x_3$ between 0.7
and 0.8 the deviation from QCD is about twenty-five percent of
the QCD result.

To exhibit the energy dependence of the \gcub\ signal we also
study the difference between the cross-section
above the cross-over point, which is roughly $x_3=.88$, and the
cross-section below this point, with and without the \gcub\ correction.
This quantity is sensitive to the \gcub\ signal because the \gcub\
correction changes sign near $x_3 = .88$.
In Fig.~5 we show the difference between the sum of the last three bins
in the $x_3$ distributions of Fig.~4 (bins 10, 11 and 12)
and the sum of the three bins below them (bins 7, 8 and 9) divided by
the sum of all six bins for different values of $\sqrt{\hat{s}}$,
$$
r_{0.88}\ =\ {
\left({d\sigma\over d\sqrt{{\hat{s}}} }\right)_{.88\leq{x_3}\leq{.95}}\
-\
\left({d\sigma\over d\sqrt{{\hat{s}}} }\right)_{.81\leq{x_3}\leq{.88}}
\over
\left({d\sigma\over d\sqrt{{\hat{s}}} }\right)_{.88\leq{x_3}\leq{.95}}\
+\
\left({d\sigma\over d\sqrt{{\hat{s}}} }\right)_{.81\leq{x_3}\leq{.88}}
}\eqn\rratio
$$
The deviations from QCD
are very large in this case, and the different energy dependence of the
two signals is clearly seen. However, including the form-factor~(\formf)
flattens the \gcub\ correction
and reduces the deviation from QCD to the point where it is smaller
than, or on the order of the statistical errors.

We have discussed three approaches to probing \gcub : one which is
limited to the region where the three jets are well separated, one
which focuses on the collinear region and one which combines these two
regions. The last two are more sensitive to the \gcub\ signal.
All the distributions we study are normalized distributions. This
should reduce some of the systematic errors in performing the
measurements.

In most of our analysis we have included the effects of a form-factor
on the deviations from QCD due to the operator \gcub. This is not
usually done in similar searches for quark substructure. Studying the
effects of a form-factor on the \gcub\ signal became necessary in the
context of dijet production, where this operator only contributes at
order $O(1/\Lambda^4)$ so that the parametrization in terms of the
effective operator breaks down at energies where the signal becomes
appreciable. In that case the form-factor greatly reduces the \gcub\
signal~[\ddz].
In the case at hand the effect of the form-factor is smaller,
but non-negligible.
Furthermore, unlike similar searches for quark substructure which use
the $p_T$ or some other energy-dependent distributions, and are
therefore sensitive to the effects of form-factors, most of the
distributions we consider here are distributions in energy-independent
quantities such as angles and energy fractions. A form-factor therefore
only reduces the deviation from QCD uniformly over the entire range, by
a factor which is equal to the energy averaged value of the form-factor,
without changing the qualitative shape of the distributions.

Throughout this discussion, we took the cut on the minimum
center-of-mass energy of the jets, $\sqrt{\hat{s}}$, to be rather low
-- 250 GeV, as used in early CDF studies~[\cdf] of three-jet events.
With the larger integrated luminosity now available at the Tevatron,
this cut can be increased in order to enhance the \gcub\ signal.
Varying this cut also allows one to distinguish between the \gcub\
signal and higher order QCD effects which have different power-law
dependences on the parton center-of-mass energy $\sqrt{\hat{s}}$.

Finally, a remark regarding the choice of $\mu$ of the coupling constant
$\alpha_s(\mu)$ is in order. In regions where the three jets are well
defined the results are not very sensitive to this choice but it
becomes relevant in regions where two of the jets approach
collinearity. The $\cos 2\varphi$ expectation value in this region is
obtained by normalizing the $\cos 2\varphi$ weighted distributions by the
non-weighted distributions so it is insensitive to the value
of the coupling. However, the $x_3$ distribution which interpolates
between the collinear and the non-collinear regions would probably be
affected. Since in this paper we only present qualitative
tree-level results this effect was not taken into account.

\vskip0.2truein
\line{ {\sl 5. Conclusions}\hfill }
\vskip0.1truein

As discussed in section~1, the operator \gcub\ cannot be reliably probed
through dijet production or inclusive jet production in hadron
colliders. Jet production in $e^+e^-$
machines cannot yield good bounds on the scale $\Lambda$ because
of the low characteristic center-of-mass energy.
This leaves three jet production in hadron colliders as the best
place to probe the gluonic operator \gcub. In this case \gcub\
contributes at leading order in $1/\Lambda^2$.

We have suggested two complementary ways of detecting the \gcub\
signal. One involves studying the region where two of the jets are
almost collinear, and the other interpolates between the collinear
region and the region where the three jets are well separated. These
approaches appear to have better analyzing power than tests that avoid
the collinear region altogether, since the deviations from QCD are more
pronounced in the collinear region. The two approaches may still be
rather challenging experimentally.

In the collinear region the \gcub\ correction, unlike the tree-level QCD
cross-section, has a non-trivial behavior under azimuthal rotations of
the collinear momenta that leave their sum unchanged. Weighting the
cross-section by the appropriate function of this azimuthal angle can
in principle allow us to separate the \gcub\ signal from QCD. The fact
that the tree-level QCD cross-section is invariant under these
azimuthal rotations follows from the helicity structure of the
four-parton QCD amplitudes at tree-level, which in turn follows from
the supersymmetric Ward identities. Thus, measuring the dependence of
the collinear three-jet
cross-section on this azimuthal angle provides a test of the
helicity structure of tree-level QCD and the SWI, as well as a method
for probing new physics in the strong-interaction sector via the
effective operator \gcub.

\vskip0.2truein
{\bf Acknowledgements}

We thank L. Randall for her collaboration at an early stage of this
work. We also thank J. D. Bjorken and D. Zeppenfeld
for useful discussions, Z. Bern for loop polynomial assistance,
and H. Haber for help with the SAGE program.
Finally, it is a pleasure to thank M. Peskin for reading the
manuscript and for insightful suggestions.

\vskip0.4truein
\vskip0.4truein

\line{ {\sl Appendix: Collinear properties of color-ordered
sub-amplitudes}\hfill }

\vskip0.1truein

In this appendix we discuss additional details of color ordered
sub-amplitudes, the helicity basis and collinear factorization.
Except for the behavior of the \gcub\ three-gluon vertex in the
collinear limits, all the results mentioned are
well known (see the review paper by Mangano and Parke~[\mangano]
and references therein) and are summarized here
for the reader's convenience.

The gluon sub-amplitudes~(\ngluon)
have the following properties:
\par\noindent
$\bullet \ A_n(1,2,\ldots,n)$ is invariant under cyclic permutations.
\par\noindent
$\bullet\ A_n(n,n-1,\ldots,1) = (-1)^n A_n(1,2,\ldots,n)$
\par\noindent
$\bullet\ A_n(1,2,3,\ldots,n)+A_n(2,1,3,\ldots,n)+\cdots
+A_n(2,3,\ldots,1,n) = 0$   \par\noindent
The last equation is due to the decoupling of the $U(1)$
gauge boson in a $U(N)$ gauge theory.
These relations reduce the number of independent sub-amplitudes that
have to be calculated.

We now turn to discuss collinear factorization in the color-ordered basis.
Equation~(\split) of section~2 can be proven from
the string representation of gauge theory tree amplitudes~[\mangano]
but here we will briefly discuss it from the point of view of
Feynman diagrams.
In the limit
when partons $i$ and $j$ become collinear, the $n$-parton
color ordered sub-amplitude with $n\geq 5$ can become singular
as the relevant propagator goes on-shell.
This only happens if the collinear partons are color-adjacent;
they have to be attached to the same three-parton
vertex to give rise to an on-shell propagator.
The singularity may be removed
by powers of the Lorentz invariant $s_{ij}$ coming from
the vertex. However, for most helicity choices of $i$ and $j$ and
$\lambda$, the QCD three-parton vertex contracted with the polarization
vectors of $i$ and $j$ (or spinors, for quarks),
behaves as $\sqrt{s_{ij}}$ at leading order
so that the singularity remains, although it is reduced
to $1/\sqrt{s_{ij}}$.
Furthermore, the coefficient of the
$\sqrt{s_{ij}}$ term in the vertex can be replaced,
up to a numerical factor,
by the polarization vector of an effective external parton with
momentum $p=k_i+k_j$ and helicity $\lambda$.
The remaining part of the diagram, contracted with the
effective parton, becomes an $(n-1)$-parton diagram, and summing over
all possible diagrams gives the $(n-1)$-parton amplitude, multiplied
by the splitting function.

We now show that no additional collinear
singularities arise from the gluon operator \gcub.
The only issue is for the quark-free $g\to gg$ singularity.
Contracting the new three gluon vertex with the polarization vectors
$\varepsilon_1$ and $\varepsilon_2$ of the external gluons 1 and 2
and using momentum
conservation and the gluons' transversality,
$\varepsilon_1\cdot k_1 = \varepsilon_2\cdot k_2 = 0$,
this vertex can be simplified to the form
$$
(k_1\cdot\varepsilon_2k_2\cdot\varepsilon_1-
k_1\cdot k_2\varepsilon_1\cdot\varepsilon_2)\ {(k_2-k_1)}^\mu \ ,
\eqn\vertex
$$
up to an overall factor which does not depend on the momenta
$k_1$, $k_2$ of the gluons\footnote\dag{
If the two gluons have opposite helicities, one can choose the reference
momentum of 1 to be $k_2$ and vice versa so that the vertex vanishes
for arbitrary $k_1$and $k_2$.}.

In the limit where $k_1$ and $k_2$ become collinear, we can write
$k_1 = zp+q$, $k_2 = (1-z)p-q$, where $z$ is a number and
$q^\mu\to 0$.
Eliminating $p$ from these relations we can express $k_1$ as a
combination of $q$ and $k_2$ and vice versa.
Using $\varepsilon_1\cdot k_1 = \varepsilon_2\cdot k_2 = 0$
the first term in eqn.~(\vertex) becomes
$$
k_1\cdot\varepsilon_2k_2\cdot\varepsilon_1\ \propto\
q\cdot\varepsilon_2q\cdot\varepsilon_1\ .
$$
But since the on-shell propagator is $1/(k_1+k_2)^2$ and
$$
(k_1+k_2)^2\ =\ 2k_1\cdot k_2\ =\ p^2 \ =\ O(q^2)\ =\ O(p\cdot q)\ ,
$$
we see that both terms in
the vertex go to zero at least as fast as $(k_1+k_2)^2$ so that
the singularity cancels.

The explicit representation of the spinor
products that we use follows ref.~[\mangano]. For two momenta with
positive energies:
$$
\spa{i}.{j}\ \equiv \ \sqrt{s_{ij}}\exp(i\phi_{ij})\ ,
\eqn\sprep
$$
where $s_{ij} = 2p_i\cdot p_j$, and
$$
\cos\phi_{ij}\ =\ (p_i^1p_j^+-p_j^1p_i^+)/\sqrt{p_i^+p_j^+s_{ij}}
\ ,\ \
\sin\phi_{ij}\ =\ (p_i^2p_j^+-p_j^2p_i^+)/\sqrt{p_i^+p_j^+s_{ij}}\ ,
$$
where $p^\pm = (p^0\pm p^3)$. If any of the momenta has negative
energy, the spinor product is calculated with minus that momentum
and then multiplied by $i$ for each negative energy momentum.
$\spb{i}.{j}$ can be calculated using
$\spa{i}.{j}\spb{j}.{i} = s_{ij}$ and the antisymmetry of the
spinor products, $\spa{i}.{j} = -\spa{j}.{i}$,
$\spb{i}.{j} = -\spb{j}.{i}$.

We collect here the tree-level splitting-functions used to infer
the \gcub\ correction to five-parton amplitudes.

The $g \to gg$ splitting functions are
$$\eqalign{
\Split_{+}(i^{+},j^{+})\ &=\ 0,\cr
\Split_{-}(i^{-},j^{-})\ &=\ 0,\cr
\Split_{-}(i^{+},j^{+})
           \ &=\ {1\over \sqrt{z (1-z)}\spa{i}.j},\cr
\Split_{+}(i^{-},j^{-})
           \ &=\ -{1\over \sqrt{z (1-z)}\spb{i}.j},\cr
\Split_{-}(i^{+},j^{-})
          \ &=\ -{z^2\over \sqrt{z (1-z)}\spb{i}.j},\cr
\Split_{+}(i^{+},j^{-})
          \ &=\ {(1-z)^2\over \sqrt{z (1-z)}\spa{i}.j},\cr
\Split_{+}(i^{-},j^{+})
          \ &=\ {z^2\over \sqrt{z (1-z)}\spa{i}.j},\cr
\Split_{-}(i^{-},j^{+})
          \ &=\ -{(1-z)^2\over \sqrt{z (1-z)}\spb{i}.j},\cr}
\eqn\gggtree
$$
the $g \to \bar{q}q$ splitting functions are
$$\eqalign{
\Split_{+}(\bar{q}^{+},q^{-})
          \ &=\ {z^{1/2}(1-z)^{3/2}
          \over \sqrt{z(1-z)}\spa{\bar{q}}.q},\cr
\Split_{+}(\bar{q}^{-},q^{+})
          \ &=\ -{z^{3/2}(1-z)^{1/2}
          \over \sqrt{z(1-z)}\spa{\bar{q}}.q},\cr
\Split_{-}(\bar{q}^{+},q^{-})
          \ &=\ {z^{3/2}(1-z)^{1/2}
          \over \sqrt{z(1-z)}\spb{\bar{q}}.q},\cr
\Split_{-}(\bar{q}^{-},q^{+})
          \ &=\ -{z^{1/2}(1-z)^{3/2}
          \over \sqrt{z(1-z)}\spb{\bar{q}}.q},\cr}
\eqn\barqqgtree
$$
and the $q \to qg$ and $\bar{q} \to g\bar{q}$ splitting functions are
$$\eqalign{
\Split_{+}(q^{-},i^{+})
          \ &=\ {z^{3/2}\over \sqrt{z(1-z)}\spa{q}.i},\cr
\Split_{+}(q^{-},i^{-})
          \ &=\ -{z^{1/2}\over \sqrt{z(1-z)}\spb{q}.i},\cr
\Split_{-}(q^{+},i^{+})
          \ &=\ {z^{1/2}\over \sqrt{z(1-z)}\spa{q}.i},\cr
\Split_{-}(q^{+},i^{-})
          \ &=\ -{z^{3/2}\over \sqrt{z(1-z)}\spb{q}.i},\cr
\Split_{-}(i^{+},\bar{q}^{+})
          \ &=\ {(1-z)^{1/2}\over \sqrt{z(1-z)}\spa{i}.{\bar{q}}},\cr
\Split_{-}(i^{-},\bar{q}^{+})
          \ &=\ -{(1-z)^{3/2}\over \sqrt{z(1-z)}\spb{i}.{\bar{q}}},\cr
\Split_{+}(i^{+},\bar{q}^{-})
          \ &=\ {(1-z)^{3/2}\over \sqrt{z(1-z)}\spa{i}.{\bar{q}}},\cr
\Split_{+}(i^{-},\bar{q}^{-})
          \ &=\ -{(1-z)^{1/2}\over \sqrt{z(1-z)}\spb{i}.{\bar{q}}}.\cr}
\eqn\qgqtree
$$
Here $z$ denotes the collinear momentum fraction, so that
$k_i = z\,p$, $k_j = (1-z)\,p$, with $p$ denoting
the sum of the collinear momenta.
Note that reversing all helicities (parity) is equivalent to exchanging
$\spa{}.{}$ and $\spb{}.{}$, and inserting a minus sign.

\vfill\eject\immediate\closeout\rfile
\centerline{{\bf References}}\bigskip\frenchspacing%
\input refs.tmp\vfill\eject\nonfrenchspacing
\vfill\eject
\vskip0.2truein

\par\noindent
{\bf Figure Captions:}
\vskip0.2truein

\par\noindent {\bf Figure 1:}
The three jet event in the parton center-of-mass frame.
The parton momenta are not drawn to scale.

\vskip0.1truein

\par\noindent {\bf Figure 2:}
Effect of \gcub\ on various distributions in three jet
event variables at $\sqrt{s}=1.8\ {\rm{TeV}}$ in the region
where the jets are well separated.
The solid line
gives the QCD prediction, and the dashed line includes the
\gcub\ correction with $\Lambda=1\ {\rm TeV}$,
with the cuts of set A (equation~(\seta)).
The error bars are the estimated statistical errors for an integrated
luminosity of 15 ${{\rm pb}}^{-1}$.

\vskip0.1truein

\par\noindent {\bf Figure 3:}
Effect of \gcub\ on various distributions of three jet
event variables at $\sqrt{s}=1.8\ {\rm{TeV}}$ in the region
where two of the jets are almost collinear.
The solid line gives the QCD prediction,
the other two lines include the
\gcub\ correction with $\Lambda=1\ {\rm TeV}$ without (dashes)
and with (dot-dashes) the form-factor
${(1+4\pi {\hat{s}}/(3\Lambda^2))}^{-2}$.
Here we impose the cuts of set B (equation~(\setb)).
The error bars are the estimated statistical errors for an integrated
luminosity of 15 ${{\rm pb}}^{-1}$.

\vskip0.1truein

\par\noindent {\bf Figure 4:}
Effect of \gcub\ on the $x_3$ distribution
at $\sqrt{s}=1.8\ {\rm{TeV}}$ in the cross-over
region.
The solid line gives the QCD prediction, the other two lines include the
\gcub\ correction with $\Lambda=1$ TeV without (dashes)
and with (dot-dashes) the form-factor
${(1+4\pi {\hat{s}}/(3\Lambda^2))}^{-2}$.
Here we impose the cuts of set C (equation~(\setc)).
The error bars are the estimated statistical errors for an integrated
luminosity of 15 ${{\rm pb}}^{-1}$.

\vskip0.1truein

\par\noindent {\bf Figure 5:}
The difference between the last three (10,11,12) bins and the
three bins below them (7,8,9) of the $x_3$ distribution of Fig. 3,
divided by the sum of the bins, as a function of $\sqrt{\hat{s}}$.
The error bars are the estimated statistical errors for an integrated
luminosity of 15 ${{\rm pb}}^{-1}$.

\bye